\documentclass[pra,superscriptaddress,showpacs,longbibliography,reprint]{revtex4-2}

\usepackage[colorlinks=true,linkcolor=blue,urlcolor=blue,citecolor=blue,pdfusetitle]{hyperref}
\usepackage{color}
\usepackage[utf8]{inputenc}
\usepackage[usenames,dvipsnames]{xcolor}
\usepackage{amsthm}
\usepackage{amsmath}
\usepackage{amssymb}
\usepackage{graphicx}
\graphicspath{{graphics/}}
\usepackage{epsfig}
\usepackage{dcolumn}
\usepackage{bm}
\usepackage{mathrsfs}
\usepackage{multirow}
\usepackage[all]{xy}
\usepackage{pbox}
\usepackage{lipsum}
\usepackage{verbatim}
\usepackage{braket}
\usepackage{dsfont}
\usepackage{array}
\usepackage{makecell}
\usepackage{tabularx}
\usepackage{isotope}
\usepackage{xr}
\usepackage{float}
\usepackage{siunitx}
\usepackage[normalem]{ulem}

\begin{document}
\raggedbottom
\title{Intermittency and metastable dark states as a resource for continuous sensing}

\newcommand{\tubingen}{Institut f\"{u}r Theoretische Physik and Center for Integrated Quantum Science and Technology,  Universit\"{a}t T\"{u}bingen, Auf der Morgenstelle 14, 72076 T\"{u}bingen, Germany.}

\author{Robert Mattes}
\affiliation{\tubingen}

\author{Igor Lesanovsky}
\affiliation{\tubingen}
\affiliation{School of Physics and Astronomy and Centre for the Mathematics and Theoretical Physics of Quantum Non-Equilibrium Systems, The University of Nottingham, Nottingham, NG7 2RD, United Kingdom.}

\author{Albert Cabot}
\affiliation{\tubingen}
\affiliation{Institute for Cross-Disciplinary Physics and Complex Systems (IFISC) UIB-CSIC, Campus Universitat Illes Balears, 07122, Palma de Mallorca, Spain.}

\date{\today}

\begin{abstract}
\noindent Quanta emitted by an open quantum system carry information about intrinsic parameters, enabling their estimation via continuous monitoring. In practice, however, only a fraction of the emitted quanta is detected, reducing the achievable sensitivity. Here, we consider few-level systems in which coherent couplings and dissipative processes compete, producing metastable dynamics characterized by emission intermittency or by the emergence of a dark state. We show that both phenomena can be beneficial for sensing but their relative performance depends strongly on the achievable detection efficiencies. Intermittent emission, marked by long alternating bright and dark periods, allows to achieve robustness with respect to inefficient detection and dephasing, whereas dark states yield significantly higher sensitivity at unit detection efficiency. Yet the latter are highly susceptible to losses. We quantify the impact of inefficient detection through the classical Fisher information of the emission record and benchmark it against the ultimate sensitivity encoded in the joint system-environment state. Finally, we demonstrate that maximum-likelihood estimators based on the observed emission record can effectively approach this sensitivity. We focus here on trapped-ion systems, however, the results extend to other quantum platforms in which similar emission dynamics can be observed.
\end{abstract}

\maketitle

\section{Introduction}

The unavoidable contact of quantum systems to their environment and the resulting loss channels are typically understood as detrimental for quantum technological applications. Recent developments in nonequilibrium physics, however, have shown that dissipation can also generate rich dynamics, including superradiant phases \cite{Bohnet2012,Ferioli2023}, phase transitions \cite{lee2012quantum,ates2012dynamical,ding2020phase,liu2024emergence}, and time-crystalline behavior \cite{kongkhambut_observation_2022,solanki2024exotic,Russo_quantum_25,Wang_boundary_25}, which can be engineered across a variety of experimental platforms. Such nonequilibrium phenomena may serve as valuable resources for metrology~\cite{montenegro_review_2024}: features such as nonequilibrium transitions \cite{macieszczak2016dynamical,ilias2022criticality,Montenegro2023,Gandia2023,Pavlov2023,cabot2025exploiting}, synchronization \cite{Xu2015,giorgi2016probing}, or time-crystalline phases \cite{Montenegro2023,cabot_continuous_2024,Iemini2024,Gribben2024,Yousefjani2025discrete,Connor2025quantum}, can enhance the sensitivity to parameters of interest.

Open quantum systems can be continuously monitored~\cite{Wiseman2009}. This allows one to estimate a parameter by gradually extracting information carried away by the environment via, e.g., photon counting or homodyne detection, rather than relying exclusively on repeated projective measurements. This approach forms the basis of \emph{continuous sensing}~\cite{Catana2012,gammelmark2013bayesian,gammelmark2014fisher,kiilerich2014estimation,macieszczak2016dynamical,kiilerich2016bayesian,Cortez2017,albarelli2018restoring,Shankar2019,Amorós-Binefa_2021,yang2023efficient,godley2023adaptive,cabot_continuous_2024,ilias2024criticality,cabot2025exploiting,mattes2025designing,Connor2025quantum,KhanTensor2025,Girotti2025estimatingquantum,lee2025timescales,jirasek2025boundary,YangQuantum2026,Radaelli2026parameterestimation,vivas2026quantum}. For setups involving cavities or waveguides, quanta dissipated into the guided or localized modes can be detected with high efficiency~\cite{cox2016deterministic,Shankar2019}. However, many platforms feature free-space spontaneous emission, where detection efficiencies typically remain low~\cite{VanDevenderEfficient2010,GhadimiScalable2017,Knollmann2026,GlicensteinPreparation2021}. Therefore, understanding how detection inefficiencies limit achievable sensitivity \cite{kiilerich2014estimation,kiilerich2016bayesian,Demkowicz_adaptive_2017,albarelli2018restoring,len2022quantum,KhanTensor2025,YangQuantum2026,Radaelli2026parameterestimation,Connor2025quantum}, and whether strategies exist that preserve high sensitivity even at low efficiencies \cite{ilias2024criticality}, is fundamentally important for continuous sensing.

\begin{figure*}[t!]
 \centering
 \includegraphics[width=\textwidth]{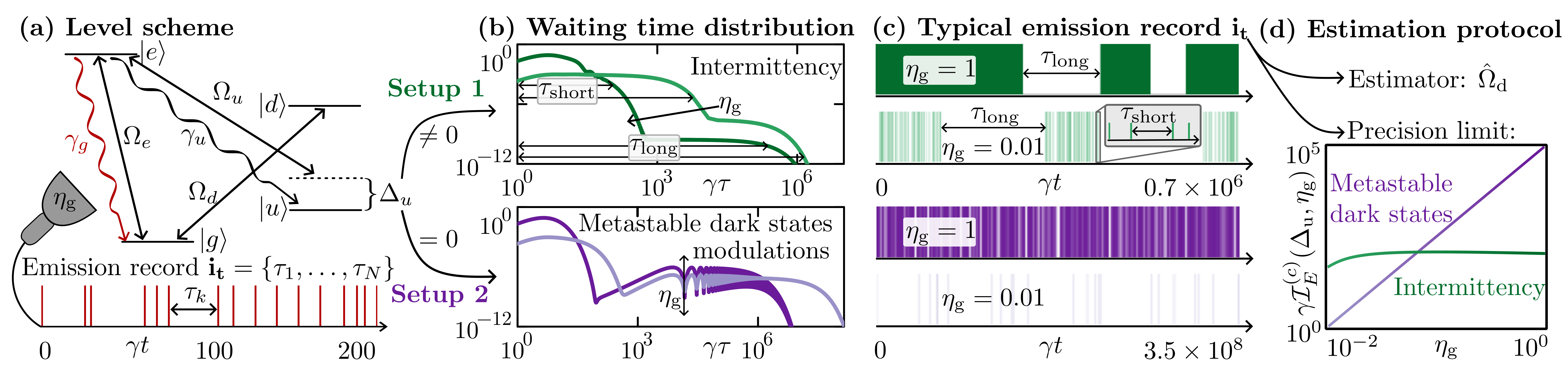}
 \caption{{\bf Continuous sensing scenarios considered in this work.} (a) Level scheme realized in trapped ion experiments. Photons emitted by the strongly fluorescent transition $\ket{e}\rightarrow\ket{g}$ are continuously monitored at a detection efficiency of $\eta_\mathrm{g}$ and the corresponding (waiting) times between emissions are collected in the emission record $\bold{i_t}$, see Sec.~\ref{sec:cont_mont_para}. By choosing the value of the detuning $\Delta_\mathrm{u}$ two qualitatively different dynamical regimes are realized characterized either by intermittency (Setup 1, $\Delta_\mathrm{u}\neq 0$) or metastable dark states (Setup 2, $\Delta_\mathrm{u}=0$).  (b) Waiting time distribution governing the statistics of the emission record $\bold{i_t}$ of Setup 1 (top) and Setup 2 (bottom) with $\eta_\mathrm{g}=0.01,1$ (bright, dark). (c) Typical emission record of Setup 1 (top) featuring intermittency and thus robustness of $\tau_{\mathrm{long}}$ for varying detection efficiencies ($\eta_\mathrm{g}=0.01,1$), with $\tau_{\mathrm{long}}$ being the long timescale appearing in the waiting time distribution of Setup 1, which is proportional to the parameter of interest $\Omega_\mathrm{d}$. The short timescale $\tau_{\mathrm{short}}$ associated to emissions within a regime of high fluorescence is indicated in the inset. Typical emission record of Setup 2 (bottom) dominated by the metastable dark state and associated modulations, which depend on $\Omega_\mathrm{d}$ and are more susceptible to inefficient detection ($\eta_\mathrm{g}=0.01,1$). (d) From the emission record $\bold{i_t}$ we find an estimate $\hat{\Omega}_\mathrm{d}$ for the unknown parameter as well as the fundamental precision limit given by the Fisher information $\mathcal{I}_\mathrm{E}^{\mathrm{(c)}}$, which crucially depends on the chosen setup and the robustness of its characteristic emission features to inefficient detection, as illustrated by the Fisher information rate [cf. Eq.~\eqref{eq:FI_rate}] sketched in the panel. Parameters: $\Omega_\mathrm{e}=2\Omega_\mathrm{u}=0.2\gamma_\mathrm{g}$, $\Omega_\mathrm{d}/\Omega_\mathrm{e}=0.0025$ and $\Delta_\mathrm{u}/\gamma=0.3(0)$ for Setup 1(2).}
 \label{fig:fig1}
\end{figure*}

Intermittent fluorescence \cite{cook1985possibility,Kimble_intermittent_1986,zoller1987quantum,schenzle1986macroscopic,kim1987quantum} offers a potential route for mitigating the limitations imposed by low detection efficiencies. This phenomenon has played a central role in trapped-ion systems where the characteristic \emph{blinking} emission pattern, formed by long periods of time with low and high fluorescence [cf. Fig.~\ref{fig:fig1}], was used to identify single quantum jumps in spite of low detection efficiencies~\cite{cook1985possibility,Kimble_intermittent_1986,zoller1987quantum,schenzle1986macroscopic,kim1987quantum,Bergquist1986Observation,Sauter_observation_1986,Nagourney_Shelved_1986,Plenio_jump_1998,Gleyzes_quantum_2007}. To this end, it was used that microscopic changes of the system state manifest in macroscopic changes of the emission record. Intermittency appears as well in many-body systems where it is rooted in collective effects~\cite{garrahan2010thermodynamics,ates2012dynamical,lee2012quantum,LesanovskyCharacterization2013,rose2016metastability,ding2020phase,liu2024emergence,CechSpaceTime2025}. Therefore, systems displaying blinking emission patterns seem natural for continuous sensing schemes involving low detection efficiencies. In fact, this idea underpinned early trapped-ion spectroscopy~\cite{Nagourney_Shelved_1986,Sauter_observation_1986,itano1987radiative,BartonLifetime2000,StaanumLifetime2004}, where the statistics of the dark periods enabled the estimation of small spontaneous emission rates.

Another intriguing phenomenon in coherently driven systems is the emergence of dark states. A system in such state effectively decouples from its environment, resulting in sustained coherences and the absence of emissions. Systems displaying such characteristic emission statistics can be engineered in various physical platforms~\cite{FleischhauerEIT2005,Donarini2019,Tolazzi2021}, in particular in trapped-ion systems, where they are exploited for cooling~\cite{morigi2000ground,LinEIT2013,LechnerEIT2016}. Dark states form the foundation of many metrology protocols~\cite{Scully1992,Vanier2005,Budker2007,Togan2011,Sedlacek2012} and, under ideal conditions, enable the saturation of ultimate sensitivity limits in continuous sensing \cite{yang2023efficient,godley2023adaptive,cabot_continuous_2024,Girotti2025estimatingquantum,lee2025timescales,jirasek2025boundary}. Weak perturbations of these systems render the dark state metastable and give rise to emissions involving, similar to intermittency, long timescales. In fact, both phenomena are different examples of metastability in open quantum systems \cite{kasha2016towards,rose2016metastability,cabot2022metastable}: the long  timescales  appearing in single quantum trajectories manifest, at the average level, as a stepwise relaxation towards the stationary state occurring over a wide range of timescales.

In this work, we demonstrate the utility of intermittency (Setup 1) and metastable dark states (Setup 2) for continuous sensing. To this end, we consider a trapped ion system [cf. Fig.~\ref{fig:fig1}(a)], which realizes both dynamical scenarios and is long-term stable, so that it can be continuously monitored over extended periods of time. Our goal is to evaluate the potential of these phenomena  for continuous sensing, focusing on realistic detection efficiencies and their impact on the optimal choice of setup. For this purpose, we study the precision limit for estimating parameters from the photodetection record given by the classical Fisher information (FI)~\cite{kiilerich2014estimation,kiilerich2015parameter,Radaelli2026parameterestimation} and compare it with the fundamental bound given by the \textit{quantum} Fisher information (QFI) associated to the system-environment state~\cite{Catana2012,gammelmark2014fisher,macieszczak2016dynamical,godley2023adaptive,yang2023efficient,cabot_continuous_2024,mattes2025designing}, which bounds any measurement making use of the information transmitted into the environment. We find that for Setup 2 the metastable dark state leads to a FI exceeding the one of Setup 1 by several orders of magnitude provided that photons can be detected with unit efficiency, see Fig.~\ref{fig:fig1}(d). However, the coherent response of the metastable dark state is more susceptible to experimental imperfections, resulting in a sensitivity that rapidly decreases with efficiency. In contrast, Setup 1 provides a lower but robust sensitivity that remains nearly optimal, i.e. reaching the fundamental limit given by the corresponding QFI. While the long-lived coherences of Setup 2 constitute a resource for continuous sensing at the upper end of realistic detection efficiencies, Setup 1 becomes favorable for low efficiencies due to its high robustness arising from intermittent emission statistics [cf. Fig.~\ref{fig:fig1}].

Finally, we remark that beyond trapped ions, our investigations are applicable to other quantum platforms implementing driven-dissipative few level systems, including NV centers \cite{yale2013all,pingault2014all}, superconducting systems \cite{zhu2014observation,zanner2022coherent}, and quantum dots \cite{KunoNonexponential2000,Efros2016}, where similar emission dynamics can be observed.

The manuscript is organized as follows. In Sec.~\ref{sec:model} we introduce the model under study, inspired by current trapped-ion platforms \cite{mallweger2024motional}. Sec.~\ref{sec:cont_mont_para} outlines the continuous monitoring scheme with inefficient photon counting and reviews key concepts in parameter estimation. Secs.~\ref{sec:blinking} and \ref{sec:modulated_blinking} present our main results on the emission dynamics and sensitivity bounds for the two configurations under study, Setup 1 and 2, respectively. In Sec.~\ref{sec:MLE} we analyze two estimation strategies based on different maximum-likelihood estimators (MLEs) and demonstrate their performance relative to the obtained sensitivity bounds. The conclusions and outlook are presented in Sec.~\ref{sec:discussion}.

\section{Trapped ion model}\label{sec:model}

For the purpose of this work we consider an ion with the level structure shown in Fig.~\ref{fig:fig1}(a). A strong separation of timescales, which forms the basis for intermittency, can be engineered by combining the strongly fluorescent transition, $\ket{g} - \ket{e}$, with a weak coherent drive on the metastable transition, $\ket{g} - \ket{d}$, i.e. $\Omega_\mathrm{d}\ll\Omega_\mathrm{e}$. In this way, the setup offers precise control over the involved timescales simply by tuning the ratio of the coherent drives. Moreover, such a level structure can be readily implemented in current experimental platforms based on $\mathrm{Sr}^{+}$ and $\mathrm{Ca}^{+}$ ions \cite{mallweger2024motional}. Here, the strongly fluorescent transition corresponds to the $S_{1/2}-P_{1/2}$ line, and $\ket{d}$ is identified with the metastable $D_{5/2}$ state. In these systems, the $P_{1/2}$ state ($\ket{e}$) can also decay into a metastable $D_{3/2}$ state, which we denote by $\ket{u}$ [cf. Fig.~\ref{fig:fig1}(a)]. For both  $\mathrm{Sr}^{+}$ and $\mathrm{Ca}^{+}$, the branching ratios are well approximated by taking the spontaneous emission rates from $\ket{e}$ to $\ket{g}$ and $\ket{u}$ to be $\gamma_\mathrm{g}=0.95\gamma$ and $\gamma_\mathrm{u}=0.05\gamma$, respectively, where $\gamma=2\pi\times 21.58$$\mathrm{MHz}$ for $\mathrm{Sr}^{+}$ ($\gamma=2\pi\times 23.05$$\mathrm{MHz}$ for $\mathrm{Ca}^{+}$). To avoid population trapping in $\ket{u}$, it is standard practice to coherently drive the transition $\ket{u}-\ket{e}$ \cite{mallweger2024motional} [cf. Fig.~\ref{fig:fig1}(a)]. As we show below, the presence of the additional level $\ket{u}$ not only enables intermittency but also leads to new effects, such as long-lived coherent modulations of the emission pattern.

The time evolution of the system's density matrix $\rho$ is governed by the master equation ($\hbar=1$):
\begin{equation}\label{eq:ME}
\dot{\rho} = -i[H,\rho]+\sum_{j} ( L_j \rho L_j^\dagger - \frac{1}{2}\{ L_j^\dagger L_j,\rho\})\, .    
\end{equation}
The Hamiltonian
\begin{equation}\label{eq:Hamiltonian}
\begin{split}
    H&= \Omega_\mathrm{e}(\ket{g}\!\bra{e} + \ket{e}\!\bra{g}) + \Omega_\mathrm{d}(\ket{g}\!\bra{d} + \ket{d}\!\bra{g})\\
    &+ \Omega_\mathrm{u}(\ket{u}\!\bra{e} + \ket{e}\!\bra{u})+\Delta_\mathrm{u} \ket{u}\!\bra{u}\, ,
\end{split}    
\end{equation}
captures the coherent part of the evolution, while the jump operators
\begin{equation}\label{eq:jump_operators_app}
    L_j=\sqrt{\gamma_j}\ket{j}\!\bra{e}\, ,\quad j = \mathrm{g}, \mathrm{u}
\end{equation}
describe the incoherent processes, where spontaneous emission from the metastable states has been neglected. We also incorporate dephasing of the long-lived state $\ket{d}$ into the model, through the jump operator $L_\mathrm{D}=\sqrt{\gamma_\mathrm{D}}\ket{d}\bra{d}$. However, we set $\gamma_\mathrm{D}=0$ unless indicated otherwise. Throughout this work, we allow for a detuning $\Delta_\mathrm{u}$ on the transition $\ket{u}-\ket{e}$, which enables the system to be operated in two qualitatively different dynamical regimes. For $\Delta_\mathrm{u}\sim \Omega_\mathrm{e,u}$ the blinking pattern is observed (Setup 1), while for $\Delta_\mathrm{u}=0$  the emission pattern is dominated by the presence of a metastable dark state (Setup 2), see Fig.~\ref{fig:fig1}. Regarding detection, we assume that only photons emitted on the $\ket{e}-\ket{g}$  transition are monitored, with an overall efficiency $\eta_\mathrm{g}$. In current trapped-ion platforms, typical detection efficiencies are at the level of a few percent~\cite{VanDevenderEfficient2010,GhadimiScalable2017,Knollmann2026}, though specialized setups can reach values that are on the order of $10\%$ or more \cite{ShuEfficient2010,wolf2020light}. We therefore place emphasis on the range $\eta_\mathrm{g}\sim 0.01 - 0.1$ throughout this work.

\section{Continuous monitoring and parameter estimation}\label{sec:cont_mont_para}

\subsection{Continuous monitoring with inefficient photon counting}

We assume that photons are recorded one by one with a detector [cf. Fig.~\ref{fig:fig1}(a)] leading under ideal conditions to the emission record $\{t_1^{L_1}, \dots, t_N^{L_N}\}$, containing apart from the jump times also the emission channel that led to the detection, which is indicated by the superscript. In practice, the finite efficiency of detection schemes means that not all emissions are recorded. Consequently, the conditional state of the system, defined with respect to the observed emission record, evolves between detections as:
\begin{equation}\label{eq:no_jump_evolution}
    \dot{\tilde{\rho}} = -i[H,\tilde{\rho}]  + \sum_{j=\mathrm{g},\mathrm{u}} ( (1-\eta_j) L_j \tilde{\rho} L_j^\dagger - \frac{1}{2}\{ L_j^\dagger L_j,\tilde{\rho}\})\, ,
\end{equation}
where $\eta_j$ denote the detection efficiencies of the corresponding emission channels \cite{Wiseman2009}, with $\eta_j\in [0,1]$ [cf. Fig.~\ref{fig:fig1}(a)].
This equation interpolates between the ideal detection case leading to a pure state evolution,  and the unmonitored case given by an evolution described by a master equation [cf. Eq.~\eqref{eq:ME}].

We assume that only the transition $\ket{e}-\ket{g}$ is monitored with an efficiency $\eta_\mathrm{g}$. This implies that the system is after every detected jump in the ground state $\ket{g}$, which is also the initial state of the evolution in between jumps [cf. Eq.~\eqref{eq:no_jump_evolution}]. The considered system therefore undergoes a so-called renewal process, where quantum jumps take the system always to the same state~\cite{kiilerich2014estimation,kiilerich2015parameter,Radaelli2026parameterestimation}.
Hence, the emission record up to a time $t$ can be entirely characterized in terms of the time differences in between consecutive detections. For $N$ detections until time $t$, the emission record is given by $\bold{i_t}=\{\tau_1, \dots, \tau_N\}$, where $\tau_k=t_{k}-t_{k-1}$ for $k\in[2,N]$ and $\tau_1=t_1$. Since we work with the fixed $t$ ensemble~\cite{Radaelli2026parameterestimation}, we also need to keep track of the time difference between the last count and $t$, denoted by $\tau_{t}=t-\sum_{k=1}^N \tau_k$. The probability density of observing the emission record $\bold{i_t}$ is:
\begin{equation}\label{eq:prob_traj}
\mathrm{P}(\bold{i_t})=\text{Tr}\big\{e^{\mathcal{L}_0\tau_{t}}\mathcal{J}_\mathrm{g} \dots \mathcal{J}_\mathrm{g}  e^{\mathcal{L}_0\tau_2}\mathcal{J}_\mathrm{g} e^{\mathcal{L}_0\tau_1}\ket{g}\!\bra{g}\big\}\, ,    
\end{equation}
where we have assumed the system is initially prepared in state $\ket{g}$. Here $\mathcal{L}_0$ corresponds to the effective Liouvillian superoperator associated to the no-jump evolution defined in Eq.~\eqref{eq:no_jump_evolution} as $\dot{\tilde{\rho}}=\mathcal{L}_0\tilde{\rho}$, with $\eta_\mathrm{u}=0$. We have also defined the jump superoperator $\mathcal{J}_\mathrm{g}\rho=\eta_\mathrm{g}L_\mathrm{g}\rho L^\dagger_\mathrm{g}$. 

For a renewal processes this probability simplifies considerably, factorizing into the probabilities of observing each individual event \cite{kiilerich2014estimation,kiilerich2015parameter,Radaelli2026parameterestimation}. These are given by the waiting time distribution (WTD), describing the probability density that two consecutive detections are separated by a time $\tau$: 
\begin{equation}\label{eq:WTD}
    W(\tau) = \eta_\mathrm{g} \gamma_\mathrm{g} \text{Tr}\big\{\ket{e}\!\bra{e} e^{\mathcal{L}_0\tau}(\ket{g}\!\bra{g})\big\}\, , 
\end{equation}
and the probability of no detection for a time interval of length $\tau$
\begin{equation}\label{eq:p_no_jump}
 P_0(\tau)=\text{Tr}\big\{ e^{\mathcal{L}_0\tau}\ket{g}\!\bra{g}\}\, .    
\end{equation}
In terms of the previous two expressions, the probability density of observing the emission record $\bold{i_t}$ is
\begin{equation}\label{eq:prob_trajectory}
\mathrm{P}(\bold{i_t})=P_0(\tau_{t}) \prod_{k=1}^{N}W(\tau_k)\, .
\end{equation}
Therefore the statistics of the emissions can be fully understood from $P_0(\tau)$ and $W(\tau)$. Moreover, notice that for long measurement times $t$, the waiting time distribution $W(\tau)$ provides the main contribution to the sensitivity of the system to changes of a parameter of interest \cite{Radaelli2026parameterestimation}. For this reason, we focus on analyzing it in the remainder of this work.

\subsection{Parameter estimation theory}

The main goal of this work is the estimation of a system parameter $\theta$ using the information contained in the counting record $\bold{i_t}$.  This is accomplished by constructing an estimator $\hat{\theta}(t)$ that returns an estimated value for the parameter of interest given $\bold{i_t}$. The stochastic nature of the counting record implies that $\hat{\theta}(t)$ is stochastic itself. Hence, its statistical  properties determine the precision by which $\theta$ can be estimated using  $\hat{\theta}(t)$ \cite{gammelmark2013bayesian,kiilerich2014estimation,Radaelli2026parameterestimation,yang2023efficient,godley2023adaptive}. In particular, the estimator is unbiased  when its average over realizations of the counting process returns the true parameter value, i.e. $\mathrm{E}[\hat{\theta}(t)]=\theta$. As figures of merit for the estimation precision, it is useful to consider both the variance of the estimator $\mathrm{Var} \,\hat{\theta}(t)=\mathrm{E[\hat{\theta}(t)^2]-E[\hat{\theta}(t)]^2}$ and the mean squared error (MSE) $\Delta \hat{\theta}(t)=\mathrm{E[(\hat{\theta}(t)-\theta)^2]}$, which coincide in the case of an unbiased estimator.

The precision of parameter estimation strongly depends on the chosen estimator. However, general bounds exist that provide universal limits on the achievable precision for a given stochastic process. In particular, the Cramér-Rao bound provides a lower bound for the variance of unbiased estimators, which for photon counting based protocols reads \cite{gammelmark2013bayesian,kiilerich2014estimation,Radaelli2026parameterestimation}: 
\begin{equation}\label{eq:CRB}
    \Delta \hat{\theta} (t)\geq \frac{1}{\mathcal{I}_\mathrm{E}^{\mathrm{(c)}}(\theta,t)}\, .
\end{equation}
Here $\mathcal{I}_\mathrm{E}^{\mathrm{(c)}}(\theta,t)$ is the FI of the counting record for a measurement time $t$, defined as:
\begin{equation}\label{eq:FI_definition}
\mathcal{I}_\mathrm{E}^{\mathrm{(c)}}(\theta,t)=\sum_{\forall \bold{i_t}} \mathrm{P}(\bold{i_t})\big[\partial_{\theta} \log \mathrm{P}(\bold{i_t}) \big]^2\, .  
\end{equation}
The sum over all possible trajectories in Eq.~(\ref{eq:FI_definition}) can be evaluated by Monte Carlo methods, see e.g. \cite{gammelmark2013bayesian,albarelli2018restoring,radaelli2024gillespie}. In our case, the product structure of the probability distribution $\mathrm{P}(\bold{i_t})$ [cf. Eq~\eqref{eq:prob_trajectory}] simplifies considerably the numerical evaluation of  $\mathcal{I}_\mathrm{E}^{\mathrm{(c)}}(\theta,t)$. For long measurement times, $\mathcal{I}_\mathrm{E}^{\mathrm{(c)}}(\theta,t)$ becomes proportional to time, and it is useful to define the long-time sensitivity rate:
\begin{equation}\label{eq:FI_rate}
\mathcal{I}^\mathrm{(c)}_\mathrm{E}(\theta)=\lim_{t\to\infty} \frac{\mathcal{I}^\mathrm{(c)}_\mathrm{E}(\theta,t)}{t}\, .
\end{equation}
For a renewal process, this quantity can be computed from the waiting time distribution using the integral formula \cite{kiilerich2014estimation,Radaelli2026parameterestimation}:
\begin{equation}\label{eq:FI_integral_formula}    
\mathcal{I}^\mathrm{(c)}_\mathrm{E}(\theta) = I_\mathrm{ss} \int_0^\infty \mathrm{d} \tau \frac{[\partial_{\theta} W(\tau)]^2}{W(\tau)}\, ,
\end{equation}
where $I_\mathrm{ss}=\eta_\mathrm{g}\gamma_\mathrm{g}\rho^{\mathrm{(ee)}}_\mathrm{ss}$ is the long-time average detection rate. As we show below,  Eq.~(\ref{eq:FI_integral_formula}) is useful to gain analytical insights on the  sensitivity of the system.

While in this work we focus on sensing using the photocounting record, a question of fundamental interest is how well this particular continuous monitoring scheme performs with respect to other possible monitoring schemes or even compared to any general measurement that uses the information contained in the joint system-environment state \cite{gammelmark2014fisher,macieszczak2016dynamical,albarelli2018restoring,yang2023efficient,godley2023adaptive,mattes2025designing}. This question can be addressed by also analyzing the QFI of the joint state of the system and all emission channels, denoted by $\mathcal{F}_\mathrm{SE}(\theta,t)$. It provides an upper bound for the FI of sensing protocols making use of the information leaked to the environment \cite{gammelmark2014fisher,macieszczak2016dynamical} (see also Appendix \ref{app:QFI} for details).  For long measurement times, this QFI is also proportional to time, and we define the long-time QFI rate as  $\mathcal{F}_\mathrm{SE}(\theta)=\lim_{t\to\infty}\mathcal{F}_\mathrm{SE}(\theta,t)/t$. Notice that $\mathcal{F}_\mathrm{SE}(\theta)$ takes into account the information encoded in {\it all} decay channels and unit efficiencies. Thereby, its comparison with the FI of the counting record with non-unit efficiency will allow us to determine how much of the ultimate sensitivity can be retrieved in experimentally accessible conditions.

In the following, we focus on the estimation of the Rabi frequency $\Omega_\mathrm{d}$, although the presented analysis can be generalized to the other system parameters. We thus drop for simplicity from now on the label $\theta$ in the FI formulas.

\section{Setup 1: Blinking system}\label{sec:blinking}

In this section, we demonstrate how long periods of high and low fluorescence can provide a robust basis for inferring parameters through continuous monitoring. To this end, we consider the trapped ion system in a regime with $\Delta_\mathrm{u}\neq 0$, $\Omega_\mathrm{u}\sim \Omega_\mathrm{e}$ and $\Omega_\mathrm{d}/\Omega_\mathrm{e}\ll 1$, where the corresponding four-level system displays the characteristic intermittent fluorescence pattern sketched in Fig.~\ref{fig:fig1}. We explicitly illustrate the robustness of the involved long timescale through a simplified analytical model, see Appendix~\ref{app:3LS}, which is based on a three-level system which displays similar physics~\cite{cook1985possibility,Kimble_intermittent_1986,zoller1987quantum,kim1987quantum,garrahan2010thermodynamics,paulino2024large}.

\subsection{Characteristic emission pattern}

\begin{figure}[t!]
 \centering
 \includegraphics[width=\columnwidth]{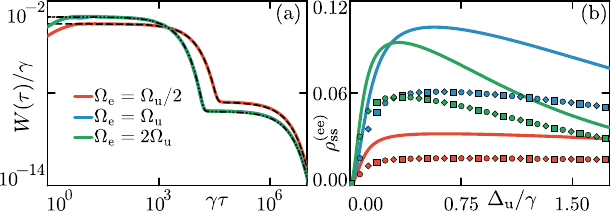}
 \caption{{\bf Emission characteristics of Setup 1.} (a) WTD featuring a separation of timescales at detection efficiency $\eta_\mathrm{g}=0.01$ and $\Omega_\mathrm{d}/\Omega_\mathrm{e}=0.001$. Black-broken lines: analytical approximations [cf. Eq.~\eqref{eq:WTD_blinking}] for the WTD. The detuning is fixed to $\Delta_\mathrm{u}/\gamma=0.3$ (green, blue) and $\Delta_\mathrm{u}/\gamma=1$ (red). (b) Stationary state population in $\ket{e}$ for $\Omega_\mathrm{d}=0$ (solid lines) and for $\Omega_\mathrm{d}/\Omega_\mathrm{e}=\{0.01,0.0025,0.001\}$ with corresponding markers $\{$squares,circles,diamonds$\}$. Color code: $\Omega_\mathrm{e}=\Omega_\mathrm{u}/2=0.1\gamma_\mathrm{g}$ (red);  $\Omega_\mathrm{e}=\Omega_\mathrm{u}=0.2\gamma_\mathrm{g}$ (blue); $\Omega_\mathrm{e}=2\Omega_\mathrm{u}=0.2\gamma_\mathrm{g}$ (green).}
 \label{fig:WT_blinking}
\end{figure}

In Fig.~\ref{fig:WT_blinking}(a), we show the WTD of Setup 1 for $\eta_\mathrm{g}=0.01$ and different values of the Rabi frequency $\Omega_\mathrm{u,e}$ and the ratio $\Omega_\mathrm{d}/\Omega_\mathrm{e}$. Further, we focus on the case in which the strongly fluorescent transition is not saturated, $\Omega_\mathrm{e} < \gamma_\mathrm{g}$. $W(\tau)$ displays in this parameter regime a strong separation of timescales, with a first exponential decay occurring at a short timescale $\tau_\mathrm{short}=1/\Gamma_1$ and a second one occurring at a much longer timescale $\tau_\mathrm{long}=1/\Gamma_2$, with $\Gamma_1 \gg \Gamma_2$. This is in fact a characteristic signature of the ensuing intermittent fluorescent dynamics depicted in Fig.~\ref{fig:fig1}(c). Here, the WTD is well-captured by the expression
\begin{equation}\label{eq:WTD_blinking}
W(\tau)=(1-p)\Gamma_1e^{-\Gamma_1 \tau}+p\Gamma_2e^{-\Gamma_2 \tau}\, ,   
\end{equation}
where $\Gamma_1\gg \Gamma_2$, and $0<p\ll 1$. While the expression for $W(\tau)$ is derived in Appendix \ref{app:3LS} for the case of the three-level blinking system, the results can be adapted to the four-level system that we consider here. In this way, we find expressions for the involved parameters, which manifest in the emission pattern associated to Setup 1 shown in the top panel of Fig.~\ref{fig:fig1}(c). $\Gamma_1$ describes the detection rate from the strongly fluorescent transition in the bright phase and $\Gamma_2$ controls the duration of the long dark periods. Intermittency [cf. Fig.~\ref{fig:fig1}(c)] is for $p \ll 1$ observed, since $p$ is approximately the inverse of the expected number of detections associated to the short timescale before an emission related to the long timescale occurs.

In particular, for $\eta_\mathrm{g}\ll 1$ and within a bright period, the strongly fluorescent transition essentially reaches its intermediate stationary state in between consecutive detections. Hence, this rate is well captured by:
\begin{equation}\label{eq:Gamma_1}
\Gamma_1=\eta_\mathrm{g}\gamma_\mathrm{g}\rho_\mathrm{ss}^\mathrm{(ee)}|_{\Omega_\mathrm{d}=0}\, ,    
\end{equation}
where $\rho_\mathrm{ss}^\mathrm{(ee)}|_{\Omega_\mathrm{d}=0}$ is the stationary population of the level $\ket{e}$ when the weak transition is not driven. The perturbation theory performed in Appendix \ref{app:3LS} further provides the second decay rate:
\begin{equation}\label{eq:Gamma_2}
\Gamma_2=\gamma \frac{\Omega_\mathrm{d}^2}{\Omega_\mathrm{e}^2}\, ,    
\end{equation}
and allows us to infer $p$:
\begin{equation}\label{eq:p}
p=\frac{\gamma^2 \Omega_\mathrm{d}^2}{4\eta_\mathrm{g}\Omega_\mathrm{e}^4}\, .    
\end{equation}
These expressions  are used to obtain the black-broken lines in Fig.~\ref{fig:WT_blinking}(a). We observe that $\Gamma_2$ is independent of detection efficiency, while the probability weight $p$ is inversely proportional to it.  Both are a consequence that, for $\Gamma_1\gg \Gamma_2$, detection inefficiency does not hinder the unambiguous identification of dark periods. In such a case, the number of total detections decreases proportionally to $\eta_\mathrm{g}$, however, the number of observed dark periods does not, leading to the dependence $p\propto\eta_\mathrm{g}^{-1}$ in Eq.~\eqref{eq:p}. Therefore, even for low detection efficiencies one can discriminate dark from bright periods, see Fig.~\ref{fig:fig1}. This allows to infer the long timescale $\tau_\mathrm{long}$, which yields information on the parameters of interest [cf. Eq.~\eqref{eq:Gamma_2}].

In order to better understand the timescale of detections within a bright period $\tau_\mathrm{short}$, and the long-time sensitivity rate, we plot in Fig.~\ref{fig:WT_blinking}(b) both $\rho_\mathrm{ss}^\mathrm{(ee)}|_{\Omega_\mathrm{d}=0}$ and the population in the excited state in the actual stationary state, $\rho_\mathrm{ss}^\mathrm{(ee)}$, for different choices of the parameter values and varying $\Delta_\mathrm{u}$. We find that $\rho_\mathrm{ss}^\mathrm{(ee)}$ is generally smaller than $\rho_\mathrm{ss}^\mathrm{(ee)}|_{\Omega_\mathrm{d}=0}$, which results from  the long periods that the system stays in the metastable $\ket{d}$ state. Further, for the considered parameter regime, $\rho_\mathrm{ss}^\mathrm{(ee)}$ does not depend significantly on $\Omega_\mathrm{d}$.

\subsection{Fisher information}

\begin{figure}[t!]
 \centering
 \includegraphics[width=\columnwidth]{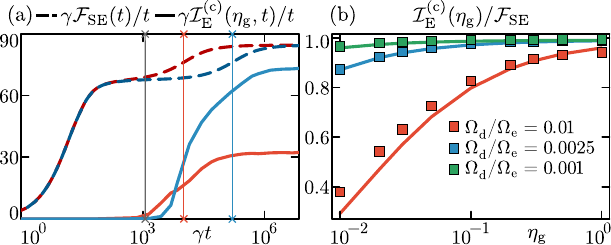}
 \caption{{\bf Sensitivity bounds for Setup 1.} (a) QFI rate (dashed) and FI rate (solid) as a function of time at detection efficiency $\eta_{\mathrm{g}}=0.01$ and $\Omega_\mathrm{d}/\Omega_\mathrm{e}=0.01,0.0025$ (red, blue). Emissions within a bright period occur on the timescale $\tau_{\mathrm{short}}=1/\Gamma_1$ [cf. Eq.~\eqref{eq:Gamma_1}] (grey vertical line), see Fig.~\ref{fig:fig1}(c). In contrast to the short timescale, $\tau_{\mathrm{long}}=1/\Gamma_2$, the timescale associated to the long dark periods, depends on $\Omega_\mathrm{d}/\Omega_\mathrm{e}=0.01,0.0025$ (red, blue vertical line) [cf. Eq.~\eqref{eq:Gamma_2}] and increases as the ratio decreases, which leads to enlarged regions of low and high fluorescence. The FI rate was calculated from a sample of $10^5$ trajectories. (b) Ratio of long-time FI rate over QFI rate as function of detection efficiency $\eta_{\mathrm{g}}$. Markers: exact results from performing numerically the integral in Eq.~(\ref{eq:FI_integral_formula}). Lines: approximation of Eq.~(\ref{eq:FI_blinking_approx}). In both cases the result is divided by $\mathcal{F}_\mathrm{SE}$ [cf. Eq.~\eqref{eq:Fisher_deformed}], which is obtained numerically. Parameters: $\Omega_\mathrm{e}=2\Omega_\mathrm{u}=0.2\gamma_\mathrm{g}$, and $\Delta_{\mathrm{u}}/\gamma=0.3$.}
 \label{fig:FI_blinking}
\end{figure}

We begin analyzing the sensitivity of Setup 1 as a function of time. To this end, we compare the ultimate sensitivity bound given by the QFI $\mathcal{F}_\mathrm{SE}$ (see Appendix \ref{app:QFI}) with the FI $\mathcal{I}_\mathrm{E}^{\mathrm{(c)}}$ associated to the emission record [cf. Eq.~\eqref{eq:FI_definition}]. The corresponding results are depicted in Fig.~\ref{fig:FI_blinking}(a). The dashed lines show the QFI, while the solid lines show the FI rate for $\eta_\mathrm{g}=0.01$. We consider two choices for the weak driving $\Omega_\mathrm{d}/\Omega_\mathrm{e}=0.01(0.0025)$, plotted in red (blue). Recall that the smaller is this ratio, the larger is the separation of timescales. We observe that, in contrast to the QFI rate, the FI rate is zero for small times before it steeply increases for times of the order of $\tau_\mathrm{short}$ (grey vertical line), which is the timescale in which photons withing a bright period are detected. Therefore, the emission record gets filled from this timescale onwards and information about the parameter of interest is collected. Both QFI rate and FI rate approach their highest values at times larger than $\tau_\mathrm{long}$ (colored vertical lines), reflecting the importance of the intermittent switching dynamics on the sensitivity to $\Omega_\mathrm{d}$. Further, we observe that the long-time QFI rate [cf. dashed lines in Fig.~\ref{fig:FI_blinking}(a)]  does not significantly depend on $\Omega_\mathrm{d}$ (see also Appendix~\ref{app:addidional_results_blinking}). In contrast, the FI at non-perfect detection efficiency, $\eta_\mathrm{g}<1$, does, as indicated by the significantly larger value for $\Omega_\mathrm{d}/\Omega_\mathrm{e}=0.0025$ compared to the one for $\Omega_\mathrm{d}/\Omega_\mathrm{e}=0.01$, see blue and red solid line in Fig.~\ref{fig:FI_blinking}(a), respectively.

We analyze in more detail the behavior of the sensitivity rates varying detection efficiency in Fig.~\ref{fig:FI_blinking}(b), in which the ratio between the long-time FI rate and QFI rate is plotted for various values of $\Omega_\mathrm{d}/\Omega_\mathrm{e}$. We observe two remarkable results. First, for $\eta_\mathrm{g}=1$ the FI rate almost saturates the QFI rate, in spite of the decay channel $\ket{e}-\ket{u}$ being completely unmonitored. Second, the long-time FI rate is very robust to detection inefficiency: for $\Omega_\mathrm{d}/\Omega_\mathrm{e}=0.001$ and $\eta_\mathrm{g}=0.01$ it still takes a value higher than $95\%$ of its maximum value. This resilience to detection inefficiency is rooted in the large separation of timescales between the fast emission processes occurring within a bright period and the slow macroscopic switching process between dark and bright periods. The resulting high contrast between these two regimes makes the slow switching dynamics, where most of the information about the parameter resides, clearly identifiable even at low detection efficiencies.

The behavior of the long-time FI rate can be better understood using Eqs. (\ref{eq:FI_integral_formula}) and (\ref{eq:WTD_blinking}), from which we obtain an approximate expression for $\mathcal{I}_\mathrm{E}^{\mathrm{(c)}}$. An important insight is that the FI rate begins to grow significantly only for times much larger than $\tau_\mathrm{short}=1/\Gamma_1$ [Fig.~\ref{fig:FI_blinking}(a)]. For such long times, $W(\tau)\approx p \Gamma_2 e^{-\Gamma_2\tau}$, which simplifies considerably the calculation of the integral in  Eq.~(\ref{eq:FI_integral_formula}). More precisely, the long-tail of the WTD begins to dominate after the time $T_\mathrm{c}$ given by:
\begin{equation}
T_c=\frac{1}{\Gamma_1-\Gamma_2}\log\bigg[\frac{(1-p)\Gamma_1}{p\Gamma_2} \bigg]\, ,
\end{equation}
at which the exponential terms in Eq.~(\ref{eq:WTD_blinking}) become equal. Then, considering only the long tail of the WTD and integrating from $\tau=T_\mathrm{c}$ on, we obtain:
\begin{equation}\label{eq:FI_blinking_approx} 
\mathcal{I}^\mathrm{(c)}_\mathrm{E}\approx I_\mathrm{ss} \int_{T_\mathrm{c}}^\infty \!\!\mathrm{d} \tau \frac{[\partial_{\Omega_\mathrm{d}}  W(\tau)]^2}{W(\tau)}\approx \frac{2\gamma^2 \gamma_\mathrm{g} \rho_\mathrm{ss}^\mathrm{(ee)}}{\Omega_\mathrm{e}^4}e^{-2\Gamma_2 T_c}\, .
\end{equation}
Since $\rho_\mathrm{ss}^{\mathrm{(ee)}}$ does not depend significantly on $\Omega_\mathrm{d}$ (see Fig.~\ref{fig:WT_blinking}) and it is independent of $\eta_\mathrm{g}$, these two parameters enter only through the exponential damping factor in Eq.~(\ref{eq:FI_blinking_approx}). The larger the separation of timescales, the smaller is $\Gamma_2 T_\mathrm{c}$. As $\eta_\mathrm{g}$ is decreased, $\Gamma_1$ decreases too, reducing the separation of timescales. For this reason, smaller values of $\Omega_\mathrm{d}/\Omega_\mathrm{e}$ translate into more resilience to detection inefficiency, as a larger value of $\Gamma_1/\Gamma_2$ ($\tau_\mathrm{long}/\tau_\mathrm{short}$) is maintained for a given $\eta_\mathrm{g}$. We use Eq.~(\ref{eq:FI_blinking_approx}) to plot the color lines in Fig.~\ref{fig:FI_blinking}(b), finding good agreement with the exact results, especially for the smallest values of $\Omega_\mathrm{d}/\Omega_\mathrm{e}$. Notice that this formula remains accurate for $\eta_\mathrm{g}=1$ too, since the long-tail of the WTD is still accurately captured by  Eq.~(\ref{eq:WTD_blinking}) (see Appendix \ref{app:3LS}), while $T_\mathrm{c}$ still captures the timescale around which the tail of the WTD becomes dominant. This is analyzed for a broader range of parameters in Appendix \ref{app:addidional_results_blinking}. Moreover, we note that by removing the exponential damping factor in Eq. (\ref{eq:FI_blinking_approx}) we obtain a simple and accurate upper bound for the FI.

Finally, we note that this scheme is also robust to dephasing acting on the metastable level. In fact, as we show in Appendix \ref{app:addidional_results_blinking}, the FI rate only begins to change and diminish significantly for $\gamma_\mathrm{D}/\gamma\sim 0.01$. For the considered trapped ion model, this means that the sensing scheme is robust to dephasing rates up to a few $\mathrm{MHz}$.

\section{Setup 2: Metastable Dark State System}\label{sec:modulated_blinking}

In this section, we consider the trapped ion system operated in the dynamical regime featuring a metastable dark state, see Fig.~\ref{fig:fig1} ($\Delta_{\mathrm{u}}= 0$), and investigate the potential of the associated emission record for parameter estimation. For this purpose, we identify in the corresponding WTD the long timescale coherent modulations as the key feature leading to an enhanced sensitivity and study its resilience for finite detection efficiencies.

\subsection{Characteristic emission pattern}\label{sec:emission_pattern_mod_blinking}

\begin{figure}[t!]
 \centering
 \includegraphics[width=\columnwidth]{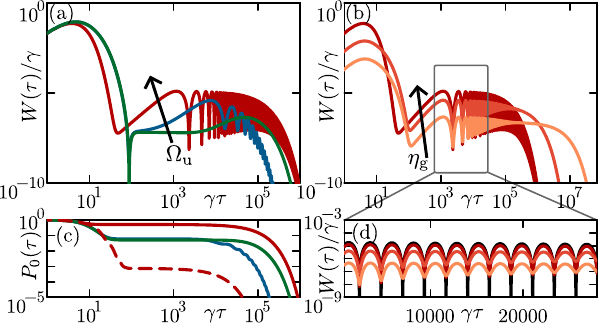}
 \caption{{\bf Emission characteristics of Setup 2.} (a) WTD featuring long timescale modulations for $\Omega_\mathrm{u}=\Omega_\mathrm{d},10\Omega_\mathrm{d},\Omega_\mathrm{e}$ (green, blue, red) and unit detection efficiency $\eta_\mathrm{g}=1$. The associated no-jump probability $P_0(\tau)$ [cf. Eq.~\eqref{eq:p_no_jump}] is depicted in (c). For comparison we show $P_0(\tau)$ for the corresponding Setup 1 ($\Delta_\mathrm{u}/\gamma=0.3$), with  $\Omega_\mathrm{u}=\Omega_\mathrm{e}$ (red dashed-line). (b) Increasing amplitude of the modulations in the WTD for increasing detection efficiencies $\eta_\mathrm{g}=0.01, 0.1, 1$ (light to dark) and $\Omega_\mathrm{u}=\Omega_\mathrm{e}$. (d) Oscillations in the WTD for intermediate times. Black line: analytical approximation from the perturbation theory [cf. Eq.~\eqref{eq:quasi_WTD_4LS}]. Parameters: $\Omega_\mathrm{e}=0.2\gamma_\mathrm{g}$, $\Omega_\mathrm{d}/\Omega_\mathrm{e}=0.01$, and $\Delta_\mathrm{u}/\gamma=0$.}
 \label{fig:WTD_osc}
\end{figure}

In Fig.~\ref{fig:WTD_osc}(a) we depict the WTD for $\eta_{\mathrm{g}}=1$ and different values of the Rabi frequency $\Omega_\mathrm{u}$. Apart from the separation of a short and a long timescale found in Setup 1 [cf. Eq.~\eqref{eq:WTD_blinking}], we also observe oscillations in the WTD on an intermediate timescale. The physical origin of these oscillations can be understood as follows. The states $\ket{g}$, $\ket{e}$, and $\ket{u}$ form the familiar $\Lambda$-level structure. Consequently, and in contrast with Setup 1, for $\Delta_{\mathrm{u}} = 0$ and $\Omega_\mathrm{d}=0$ the system possesses a dark state $\ket{D}$~\cite{morigi2000ground,FleischhauerEIT2005}, which is a superposition of $\ket{g}$ and $\ket{u}$. The Rabi drive $\Omega_\mathrm{d}$, with $\Omega_\mathrm{d} \ll \Omega_\mathrm{e,u}$, weakly couples the state $\ket{d}$ to the $\Lambda$-manifold causing $\ket{d}$ and $\ket{D}$ to hybridize. By applying perturbation theory (see Appendix~\ref{app:4LS_dark_state_manifold}) we find that the degeneracy of the hybridized states $(\ket{d}\pm\ket{D})/\sqrt{2}$ is lifted, introducing small energy shifts and finite lifetimes to the resulting states. The corresponding energy shifts manifest in the WTD as coherent modulations, with a
frequency given by [cf. Eq.~\eqref{eq:quasi_WTD_4LS}]
\begin{equation}\label{eq:freq_modulations}
    \omega=\frac{2\Omega_{\mathrm{u}}\Omega_{\mathrm{d}}}{\sqrt{\Omega_{\mathrm{e}}^2 + \Omega_{\mathrm{u}}^2}}\, .
\end{equation}
Interestingly, and as this expression clearly shows, the WTD oscillation frequency can be independently tuned by varying the Rabi frequency $\Omega_\mathrm{u}$ [see Fig.~\ref{fig:WTD_osc}(a)]. Moreover, the perturbative analysis shows that although the degeneracy within the dark state manifold is lifted at first order in $\Omega_{\mathrm{d}}$, the induced decay rate appears only at second order [cf.~Eq.~\eqref{eq:dark_second_order}]. This hierarchy leads to the separation of timescales in the WTD, where a well-defined intermediate regime emerges, characterized by coherent oscillations.

The emergent dark state has non-zero overlap with the state after a detection ($\ket{g}$). This, leads to an emission pattern characterized by long periods without any emission [cf. timescales in Fig.~\ref{fig:fig1}(c)]. This is fundamentally connected to the fact that the no-jump probability $P_0$ [cf. Eq.~\eqref{eq:p_no_jump}] features for $\Delta_{\mathrm{u}} = 0$ an extended plateau at a value that is several orders of magnitude larger than that of the corresponding Setup 1 ($\Delta_{\mathrm{u}} \neq 0$) [cf. Fig.~\ref{fig:WTD_osc}(c)]. Therefore, in contrast to Setup 1 there is no imbalance in the frequency of short and long waiting times. Hence, Setup 2 ($\Delta_{\mathrm{u}}=0$)  does not show the typical blinking pattern of many short waiting times interrupted by isolated long intervals, see Fig.~\ref{fig:fig1}(c).

After discussing the emission statistics of Setup 2 under the assumption of perfect detection efficiency  $\eta_{\mathrm{g}} = 1$ [cf. Fig.~\ref{fig:WTD_osc}(a)], we will now investigate the behavior of the WTD as $\eta_{\mathrm{g}}$ decreases. In Fig.~\ref{fig:WTD_osc}(b) we show that while the frequency of the oscillations does not change for $\eta_{\mathrm{g}}\ll 1$ and coincides with the analytical expression in Eq.~\eqref{eq:freq_modulations}, the amplitude decreases [see also Fig.~\ref{fig:WTD_osc}(d)]. Further, the long timescale increases with decreasing detection efficiency, see Appendix~\ref{Sec:WTD_app} for details. The developed physical picture can be used to understand this behavior. As the frequency of this oscillations is only related to the dark state manifold, inefficient detection does not affect this quantity.  The decreasing amplitude of the oscillations can be understood as follows: while for $\eta_{\mathrm{g}}=1$ it is very likely to observe dark periods of certain length and very unlikely to observe other dark times, for inefficient detection the length of these dark periods is blurred. Thus, the amplitude of the oscillations in the waiting time distribution decreases. On the other hand, the increased long timescale results from the fact that, if an emission occurred but was due to $\eta_{\mathrm{g}}\ll 1$ not detected, it is probable that one has to wait for the same time until another emission occurs. As previously anticipated, this is due to the finite overlap of the state after the emission with the dark state manifold. This behavior is not present in Setup 1 since the dark state manifold consists in the latter case only of the $\ket{d}$-level, which is only very weakly coupled to $\ket{g}$, see $P_0$ in Fig.~\ref{fig:WTD_osc}(c) (red dashed line).

\subsection{Fisher information}
\begin{figure}[t!]
 \centering
 \includegraphics[width=\columnwidth]{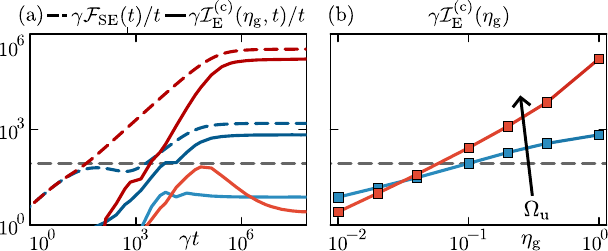}
 \caption{{\bf Sensitivity bounds for Setup 2.} (a) QFI rate $\gamma \mathcal{F}_{\mathrm{SE}}(t)/t$ (dashed) and FI rate $\gamma \mathcal{I}_\mathrm{E}^{\mathrm{(c)}}(\eta_\mathrm{g},t)/t$ (solid)  for $\Omega_{\mathrm{u}}=10\Omega_{\mathrm{d}}, \Omega_\mathrm{e}$ (blue, red) and detection efficiencies $\eta_\mathrm{g}=0.01, 1$ (light, dark). (b) Asymptotic FI rate $\gamma \mathcal{I}_\mathrm{E}^{\mathrm{(c)}}(\eta_\mathrm{g})$ for $\Omega_{\mathrm{u}}=10\Omega_{\mathrm{d}}, \Omega_\mathrm{e}$ (blue, red) as a function of detection efficiency $\eta_\mathrm{g}$. The FI rate was calculated from a sample of $\num{2e5}$ trajectories. For comparison, the asymptotic QFI rate $\gamma \mathcal{F}_{\mathrm{SE}}$ for the corresponding Setup 1 ($\Delta_{\mathrm{u}}/\gamma=0.3$) [cf. Fig.~\ref{fig:FI_blinking}(a)] is shown by the grey dashed line. Parameters: $\Omega_\mathrm{e}=0.2\gamma_\mathrm{g}$, $\Omega_\mathrm{d}/\Omega_\mathrm{e}=0.01$, and $\Delta_\mathrm{u}/\gamma=0$. }
 \label{fig:sensing_osc}
\end{figure}

We now study the sensitivity of Setup 2 ($\Delta_{\mathrm{u}} = 0$) through the associated QFI and FI rate. In the considered parameter regime, $\Omega_\mathrm{u}\sim\Omega_\mathrm{e}$ and $\Omega_\mathrm{d}\ll\Omega_\mathrm{e}$, we observe, as illustrated in Fig.~\ref{fig:sensing_osc}(a), that the ultimate precision limit given by $\mathcal{F}_{\mathrm{SE}}$ (colored dashed lines) exceeds the one of Setup 1 (grey dashed line) by several orders of magnitude. Thus, the emission statistics associated to Setup 2 [cf. Fig.~\ref{fig:WTD_osc}] host, in principle, the potential to estimate the parameter of interest $\Omega_\mathrm{d}$ to a higher precision. Although $\mathcal{I}_\mathrm{E}^\mathrm{(c)}$ for unit detection efficiency is not saturating the ultimate precision limit given by $\mathcal{F}_{\mathrm{SE}}$ (dashed lines), as shown by the dark colored solid lines in Fig.~\ref{fig:sensing_osc}(a), it still retrieves a significant amount of the QFI. Note that since photon counting is for Setup 2 optimal (see Appendix~\ref{app:optimality_photon_counting} for details), the difference can be related to the loss of information due to the unobserved decay into $\ket{u}$. Analyzing the sensitivity rates as a function of time and driving strength $\Omega_\mathrm{u}$ shows that higher sensitivities are directly connected to larger values for $\Omega_\mathrm{u}$, see Fig.~\ref{fig:sensing_osc}(a). This enhanced sensitivity is rooted in the formation of the dark state manifold for $\Delta_{\mathrm{u}} = 0$ manifesting in oscillations in the WTD, whose frequency is susceptible to changes in $\Omega_\mathrm{d}$ and which can be tuned by the Rabi drive $\Omega_\mathrm{u}$, see Eq.~\eqref{eq:freq_modulations}. Therefore, the number of observed oscillations increases with $\Omega_\mathrm{u}$, and more information about $\Omega_\mathrm{d}$ can be collected, increasing the FI rate [cf. Fig.~\ref{fig:sensing_osc}(a)].

The enhancement observed for ideal detection is more susceptible to imperfections than in Setup 1, and it diminishes significantly with decreasing efficiency, see Fig.~\ref{fig:sensing_osc}(b). Further, we observe that the slope with which the FI rate decreases with $\eta_\mathrm{g} \to 0$ generally also increases with the enhancement found for unit detection efficiency. The decreasing FI rate shows that the enhanced sensitivity in Setup 2 compared to Setup 1 is indeed not simply related to long waiting times occurring more often, which we observe for decreasing $\eta_\mathrm{g}$. By recalling Eq.~\eqref{eq:FI_integral_formula}, we explicitly see that oscillations in the WTD genuinely lead to a large gradient of the latter with respect to the parameter and thus to a high susceptibility. Therefore, the decreasing amplitude of the oscillations [cf. Fig.~\ref{fig:WTD_osc}(b)] and the concomitant smaller gradient explains the decreasing sensitivity for decreasing detection efficiency. In this sense, the emergence of the dark state manifold provides in the same way the basis for higher sensitivity as well as its lower robustness. Nevertheless, the FI rate $\gamma \mathcal{I}_\mathrm{E}^{\mathrm{(c)}}/t$ of Setup 2 is for $\eta_\mathrm{g}=0.1$  and $\Omega_\mathrm{u} \sim \Omega_\mathrm{e}$ already three times larger than the ultimate sensitivity bound of Setup 1, depicted by the grey dashed line in Fig.~\ref{fig:sensing_osc}(b).

For inefficient detection, i.e. $\eta_\mathrm{g} \ll 1 $, the optimal observation time $t'$ featuring the maximal sensitivity rate shifts to smaller times, see Fig.~\ref{fig:sensing_osc}(a). This is fundamentally connected to the increased tail of the waiting time distribution [cf. Fig.~\ref{fig:WTD_osc}(b)] being insensitive to changes in the parameter $\Omega_\mathrm{d}$, while the frequency is still sensitive to it [cf. Appendix~\ref{Sec:WTD_app}]. Further, the enhanced sensitivity relies on coherences and is thus also more susceptible to dephasing, see Appendix~\ref{Sec:osc_deph_app}. Nevertheless, one retrieves up to $70\%$ of the original FI for $\eta_\mathrm{g}=0.01$ for dephasing rates up to $\gamma_\mathrm{D}/\gamma=\num{1e-5}$.

\section{Estimation protocols}\label{sec:MLE}

In this section, we focus on particular ways to process the possible measurement results of an experiment in order to find an (optimal) estimate for the parameter that saturates the Cramér-Rao bound (CRB) in Eq.~\eqref{eq:CRB}. To this end, we present two estimation protocols, one based on a single long trajectory and the other on several short trajectories, each capable of achieving optimal performance under appropriate conditions.

\subsection{One long trajectory}
From a single measurement record $\bold{i}_\bold{t}$ up to time $t$ generated by a physical system, whose dynamics is susceptible to a parameter $\theta$, we want to infer the value of the latter. In order to explicitly connect the emission record $\bold{i}_\bold{t}$ to an estimate for the parameter, we introduce the so-called maximum likelihood estimator (MLE)
\begin{equation}\label{eq:MLE_t}
    \hat{\theta}(t)=\mathrm{argmax}_{\theta'}\,\log\mathrm{P}(\bold{i}_\bold{t}|\theta')\, .
\end{equation}
The log-likelihood function $\log\mathrm{P}(\bold{i}_\bold{t}|\theta')$ gives the (logarithm) probability of finding the exact same measurement record $\bold{i}_\bold{t}$ but for a different value of the parameter $\theta'$. Fundamentally one assumes that a certain emission-sequence is characteristic for a certain value of the parameter $\theta$ and therefore associates the MLE with the value of the parameter $\theta'$ that maximizes the probability of detecting this record, i.e. the likelihood. The fundamental theoretical precision limit for such an estimator based on a single measurement record $\bold{i}_\bold{t}$ is again given by $\mathcal{I}_\mathrm{E}^{\mathrm{(c)}}(\theta,t)$ [cf. Eq.~\eqref{eq:FI_definition}]. Interestingly, it can be shown that the mean square error (MSE) of $\hat{\theta}(t)$ follows, for large enough times and under certain regularity conditions, a normal distribution with variance given by the inverse of the FI~\cite{godley2023adaptive,Radaelli2026parameterestimation}
\begin{equation}
    (\hat{\theta}(t) - \theta )\longrightarrow\mathcal{N}(0,[\mathcal{I}_\mathrm{E}^{\mathrm{(c)}}(\theta, t)]^{-1})\, ,
\end{equation}
such that the MLE becomes asymptotically optimal.
\begin{figure}[t!]
 \centering
 \includegraphics[width=\columnwidth]{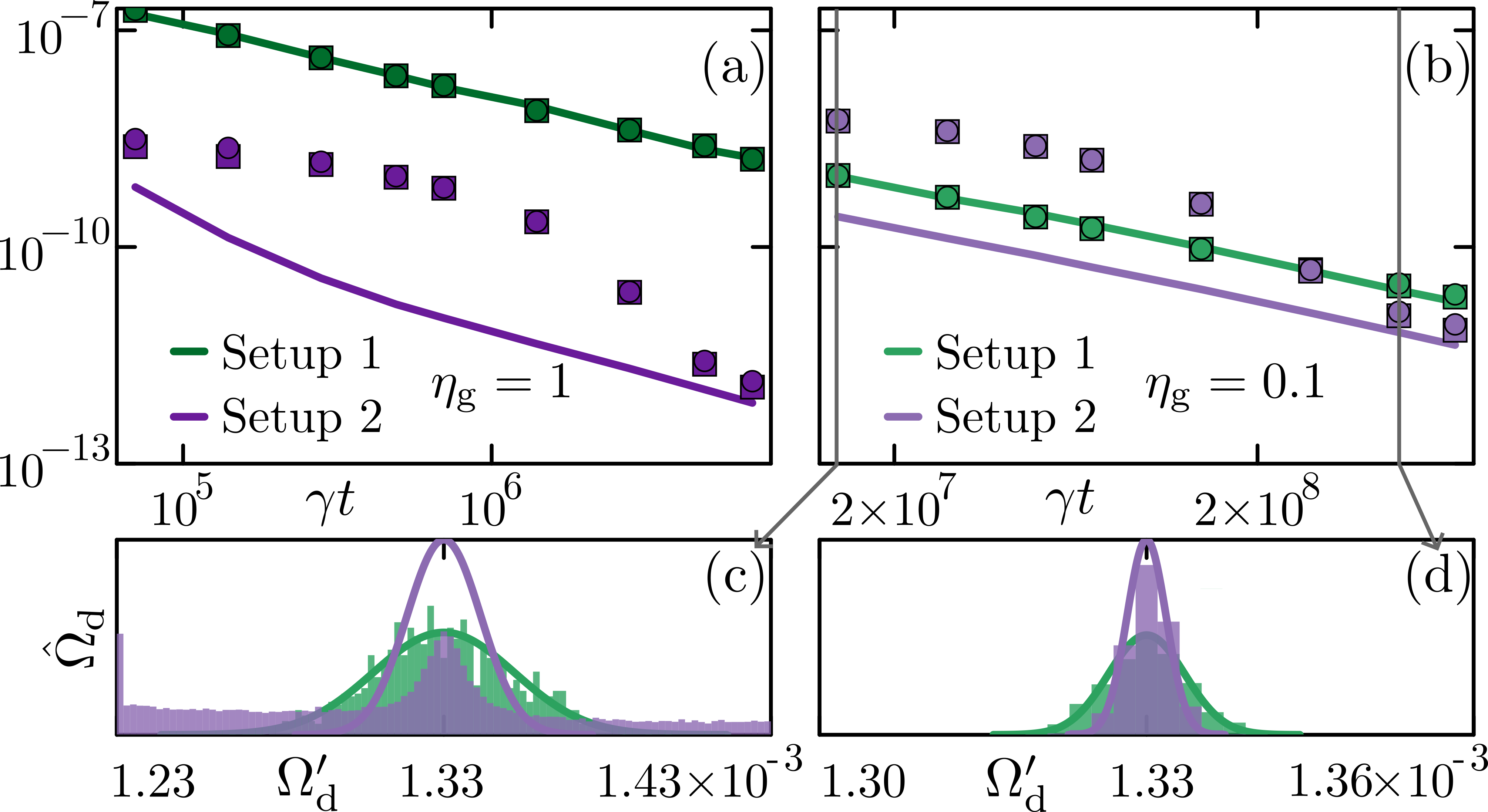}
 \caption{{\bf Maximum likelihood estimator (MLE) for one long trajectory $\bold{i}_\bold{t}$.} Evolution of the variance of the MLE $\mathrm{Var}\,\hat{\Omega}_\mathrm{d}(t)/\gamma^2$  (square), MSE $\Delta\hat{\Omega}_\mathrm{d}(t)/\gamma^2$ (circle) and the inverse of the FI $[\gamma^2 \mathcal{I}_\mathrm{E}^{\mathrm{(c)}}(\Omega_\mathrm{d},t)]^{-1}$ (solid line) for detection efficiency (a) $\eta_{\mathrm{g}}=1$  and (b) $\eta_{\mathrm{g}}=0.1$. The statistical analysis used $\num{1e3}$ trajectories for Setup 1 and $\num{5e4}$ trajectories for Setup 2. The vertical lines indicate the evaluation time of panel (c) and (d), respectively. Distribution of the estimator $\hat{\Omega}_\mathrm{d}(t)$ at $\gamma t = \num{1.4e7}$ (c)  and $\gamma t = \num{4.9e8}$ (d) for Setup 1 (green) and Setup 2 (purple). The lines correspond to normal distributions with mean equal to the true value and variance given by the inverse FI. Parameters: $\Omega_\mathrm{e}=0.2\gamma_\mathrm{g}$, $\Omega_\mathrm{d}/\Omega_\mathrm{e}=0.01$ and $\Delta_\mathrm{u}/\gamma=0.3$ (green) for Setup 1 and $\Delta_\mathrm{u}/\gamma=0$ (purple) for Setup 2. 
 }
 \label{fig:MLE_t}
\end{figure}

In Fig.~\ref{fig:MLE_t} we analyse the performance of the MLE for the two considered setups. Squares (circles) correspond to the variance (MSE) of the MLE estimation protocol, while the solid lines represent the inverse of the FI of the counting record. The variance and the MSE essentially overlap, demonstrating that for the considered observation times $t$, the MLE estimator is not significantly biased. Moreover, we observe that the CRB is for the blinking system (Setup 1) reached already for relatively short times. In contrast, for the metastable dark state system (Setup 2), one has to consider longer observation times for the variance to reach the bound. Since information about the parameter is collected by the detection of emissions and the associated waiting times, the asymptotic limit, where the CRB is reached, is related to the number of emissions. In this sense, Setup 2 features a lower emission rate due to the dark state manifold. Therefore, the CRB is for this setup only reached for very long observation times, see Fig.~\ref{fig:MLE_t}(a). On the level of the MLE this can be connected to the log-likelihood function supporting local maxima and being irregular for a small number of emission events, reflecting the physical intuition that a trajectory with a small number of emissions cannot be uniquely identified with a certain parameter. 

For decreasing detection efficiency $\eta_{\mathrm{g}} < 1$ the number of emissions per observation time decreases such that the MLE reaches the CRB in principle only for increasing observation times [cf. Fig.~\ref{fig:MLE_t}(b)]. Contrary to the blinking system (Setup 1), where for a detection efficiency of $\eta_{\mathrm{g}}=0.1$ the number of emissions is still large enough to observe asymptotic normality on the original timescale, one has to move to larger observation times for the metastable dark state system (Setup 2). Figs.~\ref{fig:MLE_t}(c)-(d) illustrate this and show that the MLE follows for both setups asymptotically a normal distribution, with a mean given by the true value and a variance given by the inverse of the FI. For the previously discussed efficiency $\eta_{\mathrm{g}}=0.01$, we expect for Setup 2 that even larger observation times are needed. Simulations are in the this case, however, computationally demanding due to the increasing tails [cf. Fig.~\ref{fig:WTD_osc}]. Nevertheless, for the Setup 1 with $\eta_{\mathrm{g}}=0.01$, the CRB is still reached for relatively short observation times, see Fig.~\ref{fig:MLE_blinking_app}.

\subsection{Multiple trajectories}

While for the blinking system the MLE in Eq.~\eqref{eq:MLE_t} reaches the CRB for a single trajectory and relatively short observation times, for the metastable dark state system very long observation times are needed. For the latter setting we show that an alternative protocol based on the collection of $N$ measurement records up to time $t$, $\bold{I}_\bold{t}^{N} = \{\bold{i}_\bold{t}^{[1]},\dots,\bold{i}_\bold{t}^{[N]}\}$, outperforms the first protocol.

Following the same ideas as in the first protocol, we assume that the collection of measurement results $\bold{I}_\bold{t}^{N}$ uniquely identifies the parameter $\theta$. With the log-likelihood function $\log \mathrm{P}(\bold{I}_\bold{t}^{N}|\theta')$ providing us with the probability of observing exactly $\bold{I}_\bold{t}^{N}$ for a different value $\theta'$, we find the MLE:
\begin{equation}\label{eq:MLE_N}
    \hat{\theta}(t,N)\!\! =\!\underset{\theta'}{\mathrm{argmax}}\log \mathrm{P}(\bold{I}_\bold{t}^{N}|\theta')\!=\!\underset{\theta'}{\mathrm{argmax}}\!\sum_{j=1}^N \log\mathrm{P}(\bold{i}_\bold{t}^{[j]}|\theta')\, .
\end{equation}
In this way one makes use of the total number of emissions within the sample $\bold{I}_\bold{t}^{N}$ such that the log-likelihood function becomes asymptotically regular with $N$, even for short times. The variance of the (unbiased) estimator $\hat{\theta}(t,N)$ is now, however, bounded by
\begin{equation}\label{eq:CRB_n}
    \Delta \hat{\theta}(t,N)\geq \frac{1}{N\mathcal{I}_\mathrm{E}^{\mathrm{(c)}}(\theta,t)}\, .
\end{equation}
\begin{figure}[t!]
 \centering
 \includegraphics[width=\columnwidth]{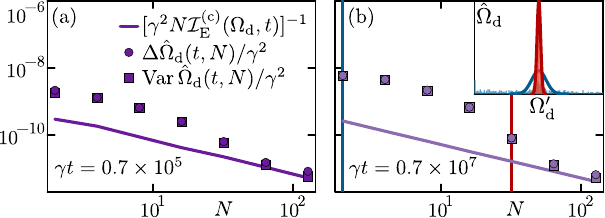}
 \caption{{\bf Maximum likelihood estimator (MLE) for Setup 2 and a collection of $N$ trajectories $\bold{I}_\bold{t}^{N}$}. Evolution of the variance of the MLE $\mathrm{Var}\,\hat{\Omega}_\mathrm{d}(t,N)/\gamma^2$  (square), MSE $\Delta\hat{\Omega}_\mathrm{d}(t,N)/\gamma^2$ (circle) and the inverse of the FI $[\gamma^2N \mathcal{I}_\mathrm{E}^{\mathrm{(c)}}(\Omega_\mathrm{d},t)]^{-1}$ (solid line) as a function of the number of measurement records $N$ used for the collection $\bold{I}_\bold{t}^{N}$. The length of the individual records is (a) $\gamma t=\num{0.7e5}$ for $\eta_{\mathrm{g}}=1$ and (b) $\gamma t=\num{0.7e7}$ for $\eta_{\mathrm{g}}=0.1$. The statistical analysis used $\num{1e3}$ collections $\bold{I}_\bold{t}^{N}$. (b) The blue and red vertical lines indicate the number of measurement records, $N$, that were used to calculate the estimator, $\hat{\theta}(t,N)$, whose distribution is illustrated in the inset. In the inset we have $\Omega'_\mathrm{d}=[\Omega_\mathrm{d}-\num{1e-4},\Omega_\mathrm{d}+\num{1e-4}]$. The lines correspond to normal distributions with mean equal to the true value and variance given by the inverse of $N$-times the FI. Parameters: $\Omega_\mathrm{e}=0.2\gamma_\mathrm{g}$, $\Omega_\mathrm{d}/\Omega_\mathrm{e}=0.01$ and $\Delta_\mathrm{u}/\gamma=0.3$.}
 \label{fig:MLE_N}
\end{figure}

In Fig.~\ref{fig:MLE_N}, we analyse the performance of the MLE defined in Eq.~\eqref{eq:MLE_N}  for the metastable dark state system (Setup 2) and short times varying the number of measurement records $N$ collected in $\bold{I}_\bold{t}^{N}$. Since the timescale shifts to larger times with decreasing detection efficiency [cf. Fig.~\ref{fig:MLE_t}] we considered $\gamma t=\num{0.7e5}$ for $\eta_{\mathrm{g}}=1$ and $\gamma t=\num{0.7e7}$ for $\eta_{\mathrm{g}}=0.1$, see Fig.~\ref{fig:MLE_N}(a) and (b), respectively. The fundamental limit given in Eq.~\eqref{eq:CRB_n}, is approached when increasing the size of the measurement collection $N$. The inset in Fig.~\ref{fig:MLE_N}(b) shows how the distribution of $\hat{\theta}(t,N)$ converges with $N$ to a normal distribution with mean given by the true value $\Omega_\mathrm{d}$ and a variance given by $[N\mathcal{I}_\mathrm{E}^{\mathrm{(c)}}(\theta,t)]^{-1}$. This illustrates the optimality of the MLE asymptotically in the sample size $N$, and that the shortcomings of the single-trajectory protocol for this setup and at short observation times appear due to insufficient sample size.

At this point, it is instructive to insert the value of $\gamma$ for, e.g., the Sr$^+$ platform to assess the actual scales implied by the sensitivity of both setups. For Setup 1, considering Fig.~\ref{fig:FI_blinking}(a) at $\gamma t=10^6$, which corresponds to  $t=0.0074$ $\mathrm{s}$, the FI has reached its long-time rate for both Rabi frequencies $\Omega_\mathrm{d}=2\pi\times 0.04(0.01)$ $\mathrm{MHz}$, shown in red (blue). For a trajectory of 1 $\mathrm{s}$ of duration (sufficient for MLE to saturate the bound), these rates yield lower  bounds on the relative estimation error of $\sqrt{\Delta \Omega_\mathrm{d}}/\Omega_\mathrm{d}=0.008(0.021)$, respectively. For Setup 2, we consider the results of Fig. \ref{fig:sensing_osc}(b) at $\eta_\mathrm{g}=0.1$, where the long-time FI rate is also reached around $\gamma t=10^6$. In this case, however, MLE requires longer times to saturate the bound [see Fig. \ref{fig:MLE_t}(b)]. In particular, for a single trajectory of duration 10 $\mathrm{s}$, or equivalently 100 trajectories of 0.1 $\mathrm{s}$, we obtain relative estimation errors of $\sqrt{\Delta \Omega_\mathrm{d}}/\Omega_\mathrm{d}=0.0015(0.0009) $ for $\Omega_\mathrm{d}=2\pi\times 0.04$ $\mathrm{MHz}$, corresponding to the blue (red) lines, respectively.

Finally, we remark that other estimation protocols are possible, which although not necessarily optimal as MLE, might be more convenient in the presence of further experimental constrains such as dark count rates.

\section{Discussion and conclusions}\label{sec:discussion}

In this work, we demonstrated the merit of intermittency (Setup 1) and metastable dark states (Setup 2) for continuous sensing under realistic detection efficiencies. For this purpose, we considered a trapped-ion system that can be implemented on current platforms \cite{mallweger2024motional} and realizes both dynamical scenarios. We analyzed the system in each configuration exhibiting either blinking ($\Delta_\mathrm{u} \neq 0$, Setup 1) or long-lived coherences due to a metastable dark state ($\Delta_\mathrm{u} = 0$, Setup 2), whereby the system is operated in closely comparable parameter regimes. For both cases we examined the sensitivity limits encoded in the Fisher information of the emission record, the upper bound set by the QFI of the system-environment state, and the performance of concrete maximum-likelihood estimation strategies.

We have found that for  Setup 1 the sensitivity is robust to inefficiency and dephasing. We have shown that this is rooted in a strong separation of timescales, and in the parameter of interest being encoded in the resilient features of the emission pattern, in particular the statistics of the switching process between dark and bright periods. Remarkably, even with 1$\%$ of detection efficiency, the achievable sensitivity can reach 95$\%$ of the QFI, and this performance can be accessed through single-trajectory MLE. With regards to Setup 2, its long-lived coherent response enhances the fundamental sensitivity by orders of magnitude under ideal conditions, yet the same dark state manifold responsible for this enhancement also reduces its robustness.  Our analysis shows that as efficiency decreases, the coherent modulation of the emission pattern dampens out, reducing the susceptibility of the statistics to changes of the parameter to be estimated. The lower emission rate found in this case further complicates single-trajectory estimation. Nevertheless, for detection efficiencies near the upper end of the experimentally accessible range, of the order of 10$\%$ \cite{wolf2020light}, this setup can still outperform the first, and this advantage remains within reach of MLE protocols.

In this way, our analysis established a clear picture and understanding of how intermittency and metastable dark states are a resource for continuous sensing and, more generally, why the optimal choice of dynamical regime depends on the detection efficiency. Our results revealed that the estimated parameter has not only to be encoded in a long timescale, but also in a feature, such as intermittency, that sustains its susceptibility for the available detection efficiencies. We note that the core of our robust continuous sensing scheme is based on intermittency, i.e. the presence of long periods of time with low and high fluorescence. Since such an emission pattern is not exclusive of trapped ion platforms~\cite{Dickson1997,NirmalFluorescence1996,KunoNonexponential2000,Efros2016}, nor of just few body systems~\cite{garrahan2010thermodynamics,ates2012dynamical,lee2012quantum,LesanovskyCharacterization2013,rose2016metastability,ding2020phase,liu2024emergence,CechSpaceTime2025}, the presented ideas and results might be of relevance for other quantum systems that feature such dynamical effects,  and parameters encoded in the macroscopic statistics of the emission events can be efficiently estimated with continuous sensing schemes. Similarly, the emergence of dark states is not unique to trapped‑ion implementations \cite{yale2013all,pingault2014all,zhu2014observation,zanner2022coherent} and can arise in many‑body settings as well~\cite{kraus2008preparation,yang2023efficient,cabot_continuous_2024}. Our analysis is therefore also relevant for these scenarios and indicates that, for the higher range of available detection efficiencies,  dark states constitute a powerful resource for continuous sensing. The exploration of these ideas in different platforms as well as in the many-body scenario, where these emission patterns can be rooted on different mechanisms, represents an intriguing venue of future research.

\section*{Acknowledgements}
We would like to thank M. Hennrich, F. Schmidt-Kaler,  V. Shankar, and R. Thomm for interesting discussions. AC acknowledges support from the Deutsche Forschungsgemeinschaft (DFG, German Research Foundation) through the Walter Benjamin programme, Grant No. 519847240 and from both the Spanish Ministerio de Ciencia, Innovación y Universidades and  Universitat de les Illes Balears through the Beatriz
Galindo programme (BG24/00134).
We acknowledge support by the state of Baden-Württemberg through bwHPC and the German Research Foundation (DFG) through grant no INST 40/575-1 FUGG (JUSTUS 2 cluster). This work was supported by the QuantERA II programme (project CoQuaDis, DFG Grant No. 532763411) that has received funding from the EU H2020 research and innovation programme under GA No. 101017733. This work is also supported by the ERC grant OPEN-2QS (Grant No. 101164443, https://doi.org/10.3030/101164443)

\appendix

\section{QFI of the system-environment joint state}\label{app:QFI}

The QFI of the joint system-environment state, $\mathcal{F}_\mathrm{SE}(\theta,t)$, provides an upper bound on the Fisher information achievable by any estimation strategy that exploits the information contained in both the system and its decay channels (the environment), including their correlations \cite{gammelmark2014fisher,macieszczak2016dynamical}. Such strategies include, but are not limited to, sequential measurements on the environment (e.g., photon counting or homodyne detection) as well as direct measurements performed on the system itself.

Under the same approximations that lead to the Lindblad master equation, the input-output formalism allows one to write an explicit (though formal) expression for this joint state (see, for instance, Refs. \cite{yang2023efficient,mattes2025designing}). Crucially, the QFI associated with this state can be computed using a significantly simpler expression that does not require constructing the full system-environment state. Instead, it relies on an auxiliary operator $\rho_{\theta_{12}}$, whose dimension matches that of the reduced system state alone \cite{Catana2012,gammelmark2014fisher,macieszczak2016dynamical}. In particular, this QFI can be obtained from:
\begin{equation}\label{eq:Fisher_deformed}
    \mathcal{F}_{\mathrm{SE}}(\theta, t) = 4 \partial_{\theta_1} \partial_{\theta_2} \log\left(\text{Tr}\{\rho_{\theta_{12}}(t)\}\right)\Big|_{\theta_1=\theta_2=\theta}\, ,
\end{equation}
where, assuming $\theta$ is a parameter of the Hamiltonian, $\rho_{\theta_{12}}$ evolves according to a two-sided master equation 
\begin{equation}\label{eq:Master_deformed}
\begin{split}
   \dot{\rho}_{\theta_{12}}=& -iH(\theta_1)\rho_{\theta_{12}}+ i\rho_{\theta_{12}}H(\theta_2)\\
   &+ \sum_{j=u,g}  ( L_j \rho_{\theta_{12}} L_j^\dagger - \frac{1}{2}\{ L_j^\dagger L_j,\rho_{\theta_{12}}\})\, ,
\end{split}   
\end{equation}
with a pure state as initial condition \cite{Catana2012,gammelmark2014fisher,macieszczak2016dynamical}.

We also note that recent works \cite{KhanTensor2025,YangQuantum2026} have proposed methods to evaluate the QFI associated with only a selected subset of the environment, for instance those decay channels that are most relevant to the metrological task or experimental setup at hand. While such approaches enable a more targeted analysis, their practical implementation is considerably more involved. For this reason, in the present work we focus exclusively on the QFI that incorporates all decay channels, i.e. $\mathcal{F}_\mathrm{SE}(\theta,t)$.

\section{Three-level blinking system}\label{app:3LS}

The physics of the four-level blinking system (Setup 1, $\Delta_\mathrm{u}\neq 0$) can be also found in the case of a simpler three-level blinking system. The latter can be well understood using perturbation theory, from which useful analytical formulas can be derived. As we have shown in the main text, these results translate to the four-level system allowing us to obtain important insights. 

Keeping this goal in mind, in this section we discuss in detail how to obtain analytical expressions for the waiting time distribution of the three-level blinking system. The blinking system that we consider is described by the following Hamiltonian:
\begin{equation}\label{eq:Hamiltonian3LS}
\begin{split}
    H&= \Omega_\mathrm{e}(\ket{g}\!\bra{e} + \ket{e}\!\bra{g}) + \Omega_\mathrm{d}(\ket{g}\!\bra{d} + \ket{d}\!\bra{g})\, ,
\end{split}    
\end{equation}
and jump operator:
\begin{equation}\label{eq:jump_operators}
L_\mathrm{g}=\sqrt{\gamma_\mathrm{g}}\ket{g}\!\bra{e}\, .
\end{equation}

\subsection{Unit detection efficiency}

Beginning with the case of unit detection efficiency, it is sufficient to consider the effective Hamiltonian
\begin{equation}
H_\mathrm{eff}=H-i\frac{\gamma_\mathrm{g}}{2}\ket{e}\!\bra{e}\, ,
\end{equation}
which we study using perturbation theory. The relevant parameter regime for us is $\Omega_\mathrm{d}\ll \Omega_\mathrm{e},\gamma_\mathrm{g}$, while we also assume the strongly fluorescent transition to be far from saturation such that $\Omega_\mathrm{e}<\gamma_\mathrm{g}/4$. In these conditions, we decompose the effective Hamiltonian as:
\begin{equation}
H_\mathrm{eff}=H_0+\Omega_\mathrm{d}V\, ,
\end{equation}
where $V=\ket{g}\bra{d} + \ket{d}\bra{g}$ is the perturbation and $H_0$ is the free effective Hamiltonian. The eigenvalues of the effective Hamiltonian are complex, and its  right and left eigenvectors satisfy:
\begin{equation}
H_\mathrm{eff}\ket{r_j}=\lambda_j\ket{r_j}\, ,\quad \bra{l_j}H_\mathrm{eff}=\lambda_j\bra{l_j}\, .
\end{equation}
Assuming $H_\mathrm{eff}$ to be diagonalizable, the eigenvectors satisfy the biorthogonality condition:
\begin{equation}
\braket{l_j|r_k}=0\,\,\text{for }j\neq k\, ,\quad \text{and } \quad \braket{l_j|r_j}\neq0\, .
\end{equation}
This allows us to use the spectral decomposition of $H_\mathrm{eff}$ to compute the waiting time distribution:
\begin{equation}\label{eq:spectral_decomp_WTD}
\begin{split}
W(\tau)&=\gamma_\mathrm{g} \big|\bra{e}e^{-iH_\mathrm{eff}\tau}\ket{g}  \big|^2\\
&=\gamma_\mathrm{g} \bigg|\sum_j\frac{\braket{l_j|g}\braket{e|r_j}}{\braket{l_j|r_j}}e^{-i\lambda_j\tau}\bigg|^2\, .
\end{split}
\end{equation}

In the following, we obtain an approximate expression for the waiting time distribution by approximating the eigenspectrum of $H_\mathrm{eff}$ using perturbation theory. The starting point is the perturbation series:
\begin{equation}
\begin{split}
\lambda_j&=\lambda_j^{(0)}+\Omega_\mathrm{d}\lambda_j^{(1)}+\Omega_\mathrm{d}^2 \lambda_j^{(2)}+\dots\\
\ket{r_j}&=\ket{r_j^{(0)}}+\Omega_\mathrm{d} \ket{r_j^{(1)}}+\Omega_\mathrm{d}^2 \ket{r_j^{(2)}}+\dots\\
\bra{l_j}&=\bra{l_j^{(0)}}+\Omega_\mathrm{d} \bra{l_j^{(1)}}+\Omega_\mathrm{d}^2 \bra{l_j^{(2)}}+\dots
\end{split}    
\end{equation}
The 0-th order corresponds to the eigenspectrum of the free effective Hamiltonian. Notice that $H_0$ is diagonalizable except for the point $\Omega_\mathrm{e}=\gamma_\mathrm{g}/4$, in which two of its eigenvectors coalesce. Away from this spectral singularity, we denote its eigenvectors and eigenvalues as:
\begin{equation}
H_0 \ket{r_j^{(0)}}=\lambda_j \ket{r_j^{(0)}}\, , \quad \bra{l_j^{(0)}} H_0=\lambda_j^{(0)} \bra{l_j^{(0)}}\, ,
\end{equation}
 and we use the labels $j\in\{d,+,-\}$ to index them. We also define
 \begin{equation}
 x=\frac{4\Omega_\mathrm{e}}{\gamma_\mathrm{g}}\, .
\end{equation} 
For $x<1$, the unperturbed eigenvalues and eigenvectors are:
\begin{equation}\label{eq:unpurturbed_eigs_blinking}
\lambda_d^{(0)}=0\, , \quad \lambda_\pm^{(0)}=-i\frac{\gamma}{4}\big(1\pm \sqrt{1-x^2}\big)\, ,
\end{equation}

\begin{equation}
\ket{r_d^{(0)}}=\ket{d}\, , \quad \bra{l_d^{(0)}}=\bra{d}\, , 
\end{equation}

\small
\begin{equation}
\begin{split}
\ket{r_+^{(0)}}=&\bigg(\frac{x}{\sqrt{1-\sqrt{1-x^2}}}\bigg)\frac{\ket{e}}{\sqrt{2}}+i\bigg(\sqrt{1-\sqrt{1-x^2}}\bigg)\frac{\ket{g}}{\sqrt{2}}\, ,\\
\bra{l_+^{(0)}}=&\bigg(\frac{x}{\sqrt{1-\sqrt{1-x^2}}}\bigg)\frac{\bra{e}}{\sqrt{2}}+i\bigg(\sqrt{1-\sqrt{1-x^2}}\bigg)\frac{\bra{g}}{\sqrt{2}}\, ,
\end{split}
\end{equation}

\begin{equation}
\begin{split}
\ket{r_-^{(0)}}=&-i\bigg(\sqrt{1-\sqrt{1-x^2}}\bigg)\frac{\ket{e}}{\sqrt{2}}
+\bigg(\frac{x}{\sqrt{1-\sqrt{1-x^2}}}\bigg)\frac{\ket{g}}{\sqrt{2}}\, ,\\
\bra{l_-^{(0)}}=&-i\bigg(\sqrt{1-\sqrt{1-x^2}}\bigg)\frac{\bra{e}}{\sqrt{2}}+\bigg(\frac{x}{\sqrt{1-\sqrt{1-x^2}}}\bigg)\frac{\bra{g}}{\sqrt{2}}\, . 
\end{split}
\end{equation}
\normalsize

From the perturbation series, using the eigenvalue equation for the 0-th order and the biorthogonality condition,  we arrive to the following expressions for the first order corrections to the eigenspectrum:
\begin{equation}
\lambda_j^{(1)}=\frac{\bra{l_j^{(0)}}V\ket{r_j^{(0)}}}{\braket{l_j^{(0)}|r_j^{(0)}}}\, ,
\end{equation}
\begin{equation}
 \ket{r_j^{(1)}}=\sum_{k\neq j}\frac{\bra{l_k^{(0)}}V\ket{r_j^{(0)}}}{\lambda_j^{(0)}-\lambda_k^{(0)}}\frac{\ket{r_k^{(0)}}}{\braket{l_k^{(0)}|r_k^{(0)}}}\, ,
\end{equation} 
\begin{equation} 
 \bra{l_j^{(1)}}=\sum_{k\neq j}\frac{\bra{l_j^{(0)}}V\ket{r_k^{(0)}}}{\lambda_j^{(0)}-\lambda_k^{(0)}}\frac{\bra{l_k^{(0)}}}{\braket{l_k^{(0)}|r_k^{(0)}}}\, .
\end{equation}
Since $V$ is off-diagonal with respect to $H_0$, it follows that:
\begin{equation}
\lambda_j^{(1)}=0\quad\forall j\, .
\end{equation}
For this reason, we also need to compute the second order correction to the eigenvalues:
\begin{equation}
\lambda_j^{(2)}=\sum_{k\neq j}\frac{1}{\lambda_j^{(0)}-\lambda_k^{(0)}}\frac{\bra{l_k^{(0)}}V\ket{r_j^{(0)}}\bra{l_j^{(0)}}V\ket{r_k^{(0)}}}{\braket{l_k^{(0)}|r_k^{(0)}}\braket{l_j^{(0)}|r_j^{(0)}}}\, .
\end{equation}

The leading perturbative corrections to the eigenspectrum can now be computed and then used to obtain the WTD. This procedure leads to a lengthy expression that is accurate in the regime of interest. A more manageable -- yet still accurate -- expression for the WTD can  be obtained by just considering the leading corrections to the eigenvalue $\lambda_d$ and eigenvectors $\ket{r_d}(\bra{l_d})$. We will focus here on this simplified expression. We then need only the following first order corrections to the right and left eigenvectors:
\begin{equation}
\begin{split}
\ket{r_d^{(1)}}&=-\frac{1}{\Omega_\mathrm{e}}\ket{e}-i\frac{\gamma_\mathrm{g}}{\Omega_\mathrm{e}^2}\ket{g}\, ,\\
\bra{l_d^{(1)}}&=-\frac{1}{\Omega_\mathrm{e}}\bra{e}-i\frac{\gamma_\mathrm{g}}{\Omega_\mathrm{e}^2}\bra{g}\, ,
\end{split}
\end{equation}
and the second order correction to the respective eigenvalue:
\begin{equation}
\lambda_d^{(2)}=-i\frac{\gamma_\mathrm{g}}{2 \Omega_\mathrm{e}^2}\, ,    
\end{equation}
Inserting these results into the spectral decomposition of the WTD and keeping only the leading terms, we obtain:
\begin{equation}
\begin{split}
W(\tau)&\approx\frac{\gamma_\mathrm{g}\Omega_\mathrm{e}^2}{\bar{\gamma}^2}\exp\bigg(-\frac{\gamma_\mathrm{g}}{2}\tau\bigg)\sinh^2\bigg(\frac{\bar{\gamma}}{4}\tau\bigg)\\
&+\frac{\gamma_\mathrm{g}^3\Omega_\mathrm{d}^4}{4\Omega_\mathrm{e}^6}\exp\bigg(-\frac{\gamma_\mathrm{g}\Omega_\mathrm{d}^2}{\Omega_\mathrm{e}^2}\tau\bigg)\\
&-\frac{2\gamma_\mathrm{g}^2\Omega_\mathrm{d}^2}{\Omega_\mathrm{e}^2\bar{\gamma}}\exp\bigg(-\bigg[\frac{\gamma_\mathrm{g}}{4}+\frac{\gamma_\mathrm{g}\Omega_\mathrm{d}^2}{2\Omega_\mathrm{e}^2}\bigg]\tau\bigg)\sinh\bigg(\frac{\bar{\gamma}}{4}\tau\bigg)\, ,
\end{split}    
\end{equation}
where we have defined:
\begin{equation}
\bar{\gamma}=\sqrt{\gamma_\mathrm{g}^2-16\Omega_\mathrm{e}^2}\, . 
\end{equation}
In Fig.~\ref{fig:3LS}(a) we compare the exact WTD with the approximate expression for $\Omega_\mathrm{d}/\Omega_\mathrm{e}=0.05$ and $\Omega_\mathrm{d}/\Omega_\mathrm{e}=0.01$. As expected, the smaller this ratio the better is the agreement. The three terms in which we have split $W(\tau)$ have the following physical interpretation. The first term corresponds to the physics in absence of the level $\ket{d}$, and it is behind the first exponential decay observed in Fig.~\ref{fig:3LS}(a). The second term describes the presence of long dark periods, whose characteristic timescale increases as $\sim \Omega_\mathrm{d}^{-2}$, and it manifests as the long-time exponential decay in Fig.~\ref{fig:3LS}(a). The third term results from the interference between the first two terms, and it is behind the dip in $W(\tau)$ observed at an intermediate timescale.

\subsection{Inefficient detection}

\begin{figure}[t!]
 \centering
 \includegraphics[width=\columnwidth]{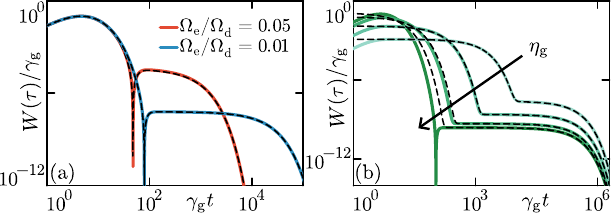}
 \caption{{\bf Waiting time distribution of the three-level blinking system.} (a) WTD for unit detection efficiency $\eta_\mathrm{g}=1$ and $\Omega_\mathrm{d}/\Omega_\mathrm{e}=0.01, 0.05$ (blue, red). Exact results: solid lines. Perturbation theory results: dashed lines. (b) WTD for $\Omega_\mathrm{d}/\Omega_\mathrm{e}=0.0025$ and varying detection efficiency $\eta_\mathrm{g}=1, 0.5, 0.1, 0.01$  (dark to bright). Black broken lines: our ansatz for the WTD, see Eq.~\eqref{eq:ansatz_pert_3LS}.}
 \label{fig:3LS}
\end{figure}

In the case of $\eta_\mathrm{g}<1$ we are not able to pursue a systematic perturbative expansion. Nevertheless, as we show below, the results of $\eta_\mathrm{g}=1$ are still useful in helping us to build an accurate ansatz for $W(\tau)$. In order to do so, we first compute numerically $W(\tau)$ and analyze its behavior as we decrease $\eta_\mathrm{g}$ and for $\Omega_\mathrm{d}/\Omega_\mathrm{e}=0.0025$, see Fig.~\ref{fig:3LS}(b). When decreasing $\eta_\mathrm{g}$ we observe the following three effects: (i) at short times $W(\tau)$ decays  exponentially with a rate that diminishes with $\eta_\mathrm{g}$; (ii) at intermediate timescales the dip resulting from interference effects disappears; (iii) the rate of the second exponential decay associated to the dark periods remains unchanged, however its relative probability weight increases with $\eta_\mathrm{g}$. 

In general, we find that for small enough (but not too small) $\eta_\mathrm{g}$ the system displays two different timescales and can be accurately described by a WTD of the form:
\begin{equation}\label{eq:ansatz_pert_3LS}
W(\tau)=(1-p)\Gamma_1  e^{-\Gamma_1 \tau}+p\Gamma_2 e^{-\Gamma_2 \tau}\, ,    
\end{equation}
where $\Gamma_1\gg\Gamma_2$, and $p$ is a probability such that $p\ll 1$. Physically, $\Gamma_1^{-1}$ describes a fast timescale related to the detecton of photons from the strongly fluorescent transition, while $\Gamma_2^{-1}$ is the slow timescale describing the dark periods. The inverse of the probability $p$ roughly corresponds to the number of detections within a bright period before switching to a dark period, such that $p\ll 1$ leads to an imbalance between emissions associated to the fast and long timescale resulting in the \emph{blinking} emission pattern. 

In the case of the three level system, we can obtain analytical expressions for these parameters. The rate $\Gamma_1$ corresponds to the detection rate within bright periods, and it can be understood as follows. Within a bright period and for small detection efficiency, the strongly fluorescent transition essentially reaches its (intermediate) stationary state in between detections. This means that the detection rate is governed by $\rho_\mathrm{ss}^{\mathrm{(ee)}}|_{\Omega_\mathrm{d}=0}$, which is the population in $\ket{e}$ in the absence of the weak driving $\Omega_\mathrm{d}$. The detection rate is then given by:
\begin{equation}
\Gamma_1=\eta_\mathrm{g}\gamma_\mathrm{g}\rho_\mathrm{ss}^{\mathrm{(ee)}}|_{\Omega_\mathrm{d}=0}=\frac{\eta_\mathrm{g}\gamma_\mathrm{g} \Omega_\mathrm{e}^2}{\frac{\gamma_\mathrm{g}^2}{4}+2\Omega_\mathrm{e}^2}\, .    
\end{equation}
As observed from the numerical results, the characteristic timescale associated to the dark periods, or long tails of $W(\tau)$ is unchanged when decreasing $\eta_\mathrm{g}$, and thus the second rate can be directly obtained from the perturbation theory results of the previous subsection:
\begin{equation}
\Gamma_2=\gamma_\mathrm{g}\frac{\Omega_\mathrm{d}^2}{\Omega_\mathrm{e}^2}\, .    
\end{equation}
The parameter $p$ can be partially extracted from the perturbation theory results too, in which we obtained analytically the coefficient in front of the long-time exponential decay. Fig.~\ref{fig:3LS}(b) shows that the probability weight of the long-time tail of $W(\tau)$ increases with decreasing $\eta_\mathrm{g}$. For $\Gamma_1\gg\Gamma_2$, this increment is well captured by a factor $\eta_\mathrm{g}^{-1}$, which when combined with the perturbative results yields:
\begin{equation}
p=\frac{\gamma_\mathrm{g}^2\Omega_\mathrm{d}^2}{4\eta_\mathrm{g}\Omega_\mathrm{e}^4}\, .    
\end{equation}
We can understand the inverse proportionality of $p$ with respect to $\eta_\mathrm{g}$ as follows. When $\Gamma_1\gg\Gamma_2$, our ability to distinguish the dark periods from the bright ones is not significantly hindered (see Fig.~\ref{fig:fig1}). This is because, in spite of the finite detection efficiency, there is still a strong separation of timescales between the characteristic timescale of detections within a bright period, $\Gamma_1^{-1}$, and the characteristic timescale governing the dark periods, $\Gamma_2^{-1}$. Therefore,  when considering a very long observation period $t\gg \Gamma_2^{-1}$, the observed number of dark periods should be the same regardless of the efficiency. Taking into account that the total number of detections decreases proportionally to $\eta_\mathrm{g}$, then the number of detections per bright period before switching to a dark period should also decrease proportionally to $\eta_\mathrm{g}$, in order for the total number of observed dark periods to be conserved. Consequently, we find that $p\propto \eta_\mathrm{g}^{-1}$ 

The resulting ansatz for the WTD is benchmarked in Fig.~\ref{fig:3LS}(b) (solid black broken lines), showing excellent agreement already for $\eta_\mathrm{g}=0.1$. Its accuracy relies on a clear separation of timescales between the characteristic detection times within a bright period and the much slower switching between bright and dark periods, i.e.,  $\Gamma_1\gg \Gamma_2$ and $p\ll1$. The same condition applies to the four-level blinking system, where one simply adapts the expressions for $\Gamma_1$, $\Gamma_2$ and $p$, as discussed in the main text.

\section{Additional results for the Blinking system (Setup 1)}\label{app:addidional_results_blinking}

\begin{figure}[t!]
 \centering
 \includegraphics[width=\columnwidth]{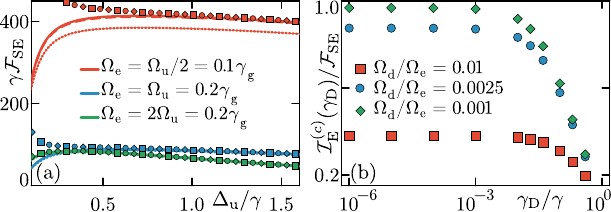}
 \caption{{\bf Additional results for the FI of the blinking system (Setup 1).} (a) Asymptotic QFI rate for $\Omega_\mathrm{d}/\Omega_\mathrm{e}=\{0.01,0.0025,0.001\}$ with corresponding markers $\{$squares,circles,diamonds$\}$. Analytical approximation for the FI, Eq.~(\ref{eq:FI_blinking_approx}),  with $\eta_\mathrm{g}=1$, and $\Omega_\mathrm{d}/\Omega_\mathrm{e}=0.001$ (solid), $\Omega_\mathrm{d}/\Omega_\mathrm{e}=0.0025$ (dashed), $\Omega_\mathrm{d}/\Omega_\mathrm{e}=0.01$ (dotted). (b)  Ratio of long-time FI rate over QFI rate varying the dephasing rate $\gamma_\mathrm{D}/\gamma$ for $\eta_\mathrm{g}=0.01$ and $\Omega_\mathrm{e}=2\Omega_\mathrm{u}=0.2\gamma_\mathrm{g}$, $\Delta/\gamma=0.3$. Notice that the QFI is the one for ideal conditions, i.e. $\gamma_\mathrm{D}=0$.}
 \label{fig:FI_blinking_appendix}
\end{figure}

In this section, we show in more detail some of the sensing capabilities of the four-level blinking system that we have introduced in the main text. 

In Fig.~\ref{fig:FI_blinking_appendix}(a) we show the long-time QFI rate as a function of the detuning and for different values of $\Omega_\mathrm{e,u,d}$. Different ratios of $\Omega_\mathrm{d}/\Omega_\mathrm{e}$ correspond to different markers, while different values of $\Omega_\mathrm{e,u}$ correspond to different colors. We observe that while the QFI rate varies significantly with $\Omega_\mathrm{e,u}$, it does not depend significantly on the precise value of $\Omega_\mathrm{d}/\Omega_\mathrm{e}$, as long as it is small enough. Moreover, notice that for not too small $\Delta_\mathrm{u}/\gamma$ the  QFI rate is well captured by the long-time FI rate with $\eta_\mathrm{g}=1$ computed with the approximate expression Eq.~(\ref{eq:FI_blinking_approx}) (see color lines). In Eq.~(\ref{eq:FI_blinking_approx}),  the ratio $\Omega_\mathrm{d}/\Omega_\mathrm{e}$ only appears within the exponential damping factor, through the term $\Gamma_2T_\mathrm{c}$, which for $\eta_\mathrm{g}=1$ and small enough $\Omega_\mathrm{d}/\Omega_\mathrm{e}$ is very small and thus $e^{-2\Gamma_2 T_\mathrm{c}}\approx 1$. From this plot we observe that the smaller $\Omega_\mathrm{d}/\Omega_\mathrm{e}$ is, the closer is the FI rate to the QFI rate, since the exponential factor tends to one.

In Fig.~\ref{fig:FI_blinking_appendix}(b) we show the ratio between the long-time FI rate varying the dephasing strength of the metastable transition and the QFI rate (without dephasing), fixing $\eta_\mathrm{g}=0.01$ and various values of $\Omega_\mathrm{d}/\Omega_\mathrm{e}$. In all cases, we observe the FI rate to be quite resilient to dephasing, only beginning to diminish for  values larger than $\gamma_\mathrm{D}/\gamma\sim0.01$.  For the considered trapped ion scheme, this values would correspond to a very large dephasing strength (of the order of $\mathrm{MHz}$). In contrast to the case of detection inefficiency, resilience to dephasing does not seem to depend significantly on the ratio $\Omega_\mathrm{d}/\Omega_\mathrm{e}$.

\section{Additional results for the metastable dark state system (Setup 2)}
\label{app:4LS_dark_state_manifold}
\subsection{Unit detection efficiency in both channels}

Analog to the blinking system (Setup 1) we make use of a non-Hermitian perturbation theory in order to find analytical expressions for the characteristic features of the WTD observed before, see Sec.~\ref{sec:emission_pattern_mod_blinking}. In contrast to the case studied in Appendix~\ref{app:3LS} we explicitly have to consider all decay channels and assume $\eta_{\mathrm{g}} = \eta_{\mathrm{u}} = 1 $, such that the state remains pure during the evolution between two jumps and a non-Hermitian perturbation theory is applicable. Therefore, the spectral decomposition given in Eq.~\eqref{eq:spectral_decomp_WTD} does not immediately coincide with the waiting time distribution of Sec.~\ref{sec:emission_pattern_mod_blinking}, since the latter considers the case $\eta_{\mathrm{u}} = 0$. In order to distinguish the resulting expression from the WTD we denote it by $W'(\tau)$ and define it as $W'(\tau)= \bar{\gamma}_\eta \big|\bra{e}e^{-iH_\mathrm{eff}\tau}\ket{g}  \big|^2$, where $\bar{\gamma}_\eta = (\gamma_\mathrm{g}\eta_\mathrm{g}+\gamma_\mathrm{u}\eta_\mathrm{u})$. As we will show in the following, the features we are interested in remain preserved for $\eta_{\mathrm{u}} \to 0$ and thus allows us to translate them from $W'$ to $W$ [cf. Eq.~\eqref{eq:WTD}].

The starting point is again the effective Hamiltonian
\begin{equation}
    H_\mathrm{eff}=H-i\sum_{j=\mathrm{g},\mathrm{u}}\frac{\gamma_j}{2}L_j^\dagger L_j = H_0-i\frac{\gamma}{2}\ket{e}\!\bra{e}\, ,
\end{equation}
with $H$ given in Eq.~\eqref{eq:Hamiltonian} and $\gamma = \gamma_{\mathrm{g}}+\gamma_{\mathrm{u}}$. To zero order in $\Omega_\mathrm{d}$, we find the following eigenvalues with associated eigenvectors
\begin{equation}
    \begin{split}
    \lambda_1^{(0)} = 0:&\quad  \ket{r_1^{(0)}} = \ket{d}\, \quad  \bra{l_1^{(0)}} = \bra{d}\, , \\\
    \lambda_2^{(0)} = 0:&\quad   \ket{r_2^{(0)}} = \ket{D} = (\Omega_{\mathrm{u}}\ket{g}-\Omega_{\mathrm{e}}\ket{u})/\Omega\, , \\\
    &\quad\bra{l_2^{(0)}} = \bra{D} = (\Omega_{\mathrm{u}}\bra{g}-\Omega_{\mathrm{e}}\bra{u})/\Omega\, , \\\
    \lambda_3^{(0)} = \lambda_+:& \quad  \ket{r_3^{(0)}} = \Omega_{\mathrm{e}}\ket{g}+\lambda_+\ket{e}+\Omega_{\mathrm{u}}\ket{u}\, , \\\
    & \quad  \bra{l_3^{(0)}} = \Omega_{\mathrm{e}}\bra{g}+\lambda_+\bra{e}+\Omega_{\mathrm{u}}\bra{u}\, , \\\
    \lambda_4^{(0)} = \lambda_-:&\quad   \ket{r_4^{(0)}} = \Omega_{\mathrm{e}}\ket{g}+\lambda_-\ket{e}+\Omega_{\mathrm{u}}\ket{u}\, ,\\\
    &\quad   \bra{l_4^{(0)}} = \Omega_\mathrm{e}\bra{g}+\lambda_-\bra{e}+\Omega_{\mathrm{u}}\bra{u}\, ,
    \end{split}
\end{equation}
where $\Omega = \sqrt{\Omega_{\mathrm{e}}^2 + \Omega_{\mathrm{u}}^2}$ and $\lambda_\pm$ as in \eqref{eq:unpurturbed_eigs_blinking}, with $x = 4\Omega/\gamma$. Since we are interested in the regime $\Omega_{\mathrm{u}}\sim\Omega_{\mathrm{e}}$, we focus on the case $x>1$.

Note that the first two eigenvalues are degenerated. Thus, we construct new eigenvectors in the corresponding subspace for which the perturbation $\Omega_{\mathrm{d}}(\ket{g}\bra{d} + \ket{d}\bra{g}) = \Omega_{\mathrm{d}} V$ becomes diagonal. With these new eigenvectors, e.g. $\ket{\tilde{r}_{1,2}^{(0)}} = (\ket{d}\pm \ket{D})/\sqrt{2}$ and $\ket{\tilde{r}_{3,4}^{(0)}}=\ket{r_{3,4}^{(0)}}$, we find the first order correction
\begin{equation}
    \begin{split}
        \tilde{\lambda}_1^{(1)} &= \frac{\braket{\tilde{l}_{1}^{(0)}|(\ket{g}\bra{d} + \ket{d}\bra{g})|\tilde{r}_{1}^{(0)}}}{\braket{\tilde{l}_{1}^{(0)}|\tilde{r}_{1}^{(0)}}} = \frac{\Omega_{\mathrm{u}}}{\Omega}\, , \\\
        \tilde{\lambda}_2^{(1)} &= \frac{\braket{\tilde{l}_{2}^{(0)}|(\ket{g}\bra{d} + \ket{d}\bra{g})|\tilde{r}_{2}^{(0)}}}{\braket{\tilde{l}_{2}^{(0)}|\tilde{r}_{2}^{(0)}}} = -\frac{\Omega_{\mathrm{u}}}{\Omega}\, ,       
    \end{split}
\end{equation}
such that the degeneracy is lifted at first order in $\Omega_{\mathrm{d}}$. We emphasize that this is contrary to the blinking system, where the first non-vanishing contribution of the perturbation is second order in $\Omega_{\mathrm{d}}$. The second order correction reads
\begin{equation}
\begin{split}\label{eq:dark_second_order}
        \tilde{\lambda}_{1/2}^{(2)} &= \left( \frac{|V_{3,1/2}|^2}{-\lambda_+\braket{\tilde{l}_{3}^{(0)}|\tilde{r}_{3}^{(0)}}} + \frac{|V_{4,1/2}|^2}{-\lambda_-\braket{\tilde{l}_{4}^{(0)}|\tilde{r}_{4}^{(0)}}} \right) \\\
        &= -\frac{\Omega_{\mathrm{e}}^2}{2}\underbrace{ \frac{2i\gamma-i\gamma^3/(8\Omega^2)}{ 4\Omega^4+(-i\gamma/2)^2\Omega^2}}_{\chi}\, ,
\end{split}
\end{equation}
where we used that $V_{jk} = \braket{\tilde{l}_{j}^{(0)}|(\ket{g}\!\bra{d} + \ket{d}\!\bra{g}) |\tilde{r}_{k}^{(0)}} = \Omega_\mathrm{e}/ \sqrt{2}$, for $k=1,2$ and $j=3,4$. Since the second order corrections $\tilde{\lambda}_{1/2}^{(2)}$ are purely imaginary they only introduce a finite lifetime without changing the frequency. To first order in $\Omega_{\mathrm{d}}$ we find the right eigenvectors
\begin{equation}
\begin{split}
    \ket{\tilde{r}_{1/2}} &= \ket{\tilde{r}_{1/2}^{(0)}} + \Omega_{\mathrm{d}} \sum_{m=3,4} \frac{V_{m,1/2}}{\lambda_m^{(0)}\braket{\tilde{l}_{m}^{(0)}|\tilde{r}_{m}^{(0)}}} \Big( -\ket{\tilde{r}_{m}^{(0)}}\\\
    &+ \frac{V_{2/1,m}}{\lambda_{2/1}^{(1)} - \lambda_{1/2}^{(1)}}  \ket{\tilde{r}_{2/1}^{(0)}}\Big)\, ,
\end{split}
\end{equation}
and analog the left eigenvectors. While for the other two eigenvectors the correction to their eigenvalue in first order vanishes, we find the eigenvectors to first order
\begin{equation}
       \ket{\tilde{r}_{3/4}^{(1)}} = \sum_{m=1,2} \frac{V_{m,3/4}}{\lambda_{3/4}^{(0)}\braket{\tilde{l}_{m}^{(0)}|\tilde{r}_{m}^{(0)}}} \ket{\tilde{r}_{m}^{(0)}} = \frac{\Omega_{\mathrm{e}}}{\lambda_{3/4}^{(0)}} \ket{d}\, ,
\end{equation}
and the second order correction for the eigenvalues
\begin{equation}
    \frac{\braket{\tilde{l}_{3/4}^{(0)}|(\ket{g}\bra{d} + \ket{d}\bra{g})|\tilde{r}_{3/4}^{(1)}}}{\braket{\tilde{l}_{3/4}^{(0)}|\tilde{r}_{3/4}^{(0)}}} = \frac{\Omega_{\mathrm{e}}^2}{\lambda_{3/4}^{(0)}\braket{\tilde{l}_{3/4}^{(0)}|\tilde{r}_{3/4}^{(0)}}}\, .
\end{equation}

\begin{figure}[t!]
 \centering
 \includegraphics[width=\columnwidth]{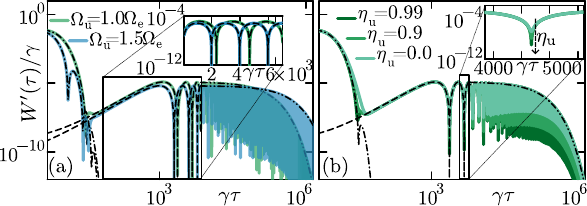}
 \caption{{\bf Modulations in the evolution between jumps for Setup 2.} (a) Numerical (solid lines) and analytical (broken lines) results for the evolution between two jumps $W'(\tau)$, i.e.  $\eta_\mathrm{g}=\eta_\mathrm{u}=1$, with $\Omega_{\mathrm{u}} = \Omega_{\mathrm{e}}, 1.5\Omega_{\mathrm{e}}$ (green, blue). (b) Numerical (solid lines) and analytical (broken lines) results for the evolution between two jumps $W'(\tau)$ for $\Omega_{\mathrm{u}} = \Omega_{\mathrm{e}}$, $\eta_\mathrm{g}=1$ and $\eta_\mathrm{u}=0.99, 0.9, 0$.
 Broken lines: analytical approximation dominating for short times [dash-dot; Eq.~\eqref{eq:short_time_no_jump}], long times  [dash; Eq.~\eqref{eq:quasi_WTD_4LS}], and exponential decay in the long-time limit [dash-dot-dot; Eq.~\eqref{eq:quasi_WTD_4LS}]. Parameters: $\Omega_\mathrm{e}=0.2\gamma_\mathrm{g}$, $\Omega_\mathrm{d}/\Omega_\mathrm{e}=0.01$, and $\Delta_\mathrm{u}/\gamma=0$ (Setup 2). }
 \label{fig:4LS_analytical}
\end{figure}
For the inner products appearing in the expression for the waiting time distribution [cf. Eq.~\eqref{eq:spectral_decomp_WTD}] we therefore find
\begin{equation}\label{eq:inner_products_mod_blinking}
    \begin{split}
    \tilde{\lambda}_1:&\quad  \braket{\tilde{l}_{1}|g} = \frac{\Omega_{\mathrm{u}}}{\sqrt{2\Omega^2}} + \frac{\Omega_{\mathrm{e}}\Omega_{\mathrm{d}}}{\sqrt{2}}\cdot\left(-3\chi\Omega_{\mathrm{e}}/4 \right)\, ,\\\
    &\quad  \braket{e|\tilde{r}_{1}}=-\frac{\Omega_{\mathrm{e}}\Omega_{\mathrm{d}}}{\sqrt{2}\Omega^2}\, ,\\\
    \tilde{\lambda}_2:&\quad   \braket{\tilde{l}_{2}|g} =-\frac{\Omega_{\mathrm{u}}}{\sqrt{2\Omega^2}} + \frac{\Omega_{\mathrm{e}}\Omega_{\mathrm{d}}}{\sqrt{2}}\left(-3\chi\Omega_{\mathrm{e}}/4\right)\, ,\\\
    &\quad  \braket{e|\tilde{r}_{2}}=-\frac{\Omega_{\mathrm{e}}\Omega_{\mathrm{d}}}{\sqrt{2}\Omega^2}\, ,\\\   
    \tilde{\lambda}_3:& \quad  \braket{\tilde{l}_{3}|g} =\Omega_{\mathrm{e}}\, ,  \quad  \braket{e|\tilde{r}_{3}}=\lambda_3^{(0)}\, ,\\\
    \tilde{\lambda}_4:&\quad   \braket{\tilde{l}_{4}|g} =\Omega_{\mathrm{e}}\, , \quad   \braket{e|\tilde{r}_{4}}=\lambda_4^{(0)}\,  .  
    \end{split}
\end{equation}
For the considered regime the contribution of the bright subspace decays with $\lambda_\pm$ and is thus zero order in $\Omega_{\mathrm{d}}$ such that their influence is limited to short times and follows
\begin{equation}\label{eq:short_time_no_jump}
\begin{split}
    W'(\tau) &\approx \bar{\gamma}_\eta \Big|\frac{\lambda_+\Omega_{\mathrm{e}}e^{-i\lambda_+ \tau}}{\Omega_{\mathrm{e}}^2+\Omega_{\mathrm{u}}^2+\lambda_+^2+(\Omega_{\mathrm{d}}^2\Omega_{\mathrm{e}}^2)/\lambda_+^2} \\\
    &+\frac{\lambda_-\Omega_{\mathrm{e}}e^{-i\lambda_- \tau}}{\Omega_{\mathrm{e}}^2+\Omega_{\mathrm{u}}^2+\lambda_-^2+(\Omega_{\mathrm{d}}^2\Omega_{\mathrm{e}}^2)/\lambda_-^2}\Big|^2\, .
    \end{split}
\end{equation}
As depicted in Fig.~\ref{fig:4LS_analytical}, the numerical solution for the evolution between two jumps (solid line) perfectly coincides for short times with the analytical expression given above [cf. Eq.~\eqref{eq:short_time_no_jump}] (dashed-dotted line). Beyond this short timescale associated to the bright subspace the evolution between two jumps is for unit detection efficiency in both channels well-captured by
\begin{equation}\label{eq:quasi_WTD_4LS}
\begin{split}
W'(\tau) &\approx \bar{\gamma}_\eta \Big|e^{-i\Omega_{\mathrm{d}}^2\tilde{\lambda}_{1/2}^{(2)}\tau}\Big|^2 \Big|\Big[\frac{\Omega_{\mathrm{e}}\Omega_{\mathrm{d}}\Omega_{\mathrm{u}}}{\Omega^3}\sin(\Omega_{\mathrm{d}}\tilde{\lambda}_{1/2}^{(1)} \tau) \\\
&+\frac{3 \chi}{4}\frac{\Omega_{\mathrm{e}}^3\Omega_{\mathrm{d}}^2}{\Omega^2} \cos(\Omega_{\mathrm{d}}\tilde{\lambda}_{1/2}^{(1)} \tau)\Big]\Big|^2\, , 
\end{split}
\end{equation}
as illustrated by the dashed line in Fig.~\ref{fig:4LS_analytical}. This corresponds to the inner products related to the dark subspace [cf. $\tilde{\lambda}_{1/2}$ in Eq.~\eqref{eq:inner_products_mod_blinking}]. While here the first order of the perturbation theory leads to oscillations, the second order provides their decay [cf.  broken line (dash-dot-dot) in Fig.~\ref{fig:4LS_analytical}]. Hence, the evolution between two jumps shows sustained oscillations over a large timescale. By taking also interference terms between the inner products associated to the dark and bright subspace into account one interpolates between them and perfectly recovers the numerical solution for all times.

\subsection{Waiting time distribution}\label{Sec:WTD_app}

\begin{figure}[t!]
 \centering
 \includegraphics[width=\columnwidth]{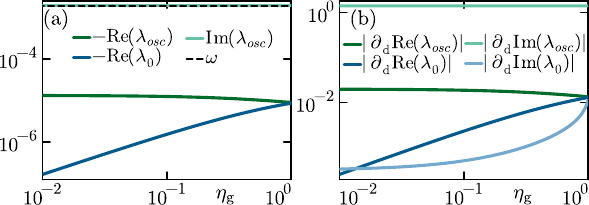}
 \caption{{\bf Spectral decomposition of $\mathcal{L}_0$ for Setup 2.} (a) Eigenvalues of the no-jump generator $\mathcal{L}_0$ for varying detection efficiency $\eta_\mathrm{g}$. Blue: overall decay rate. Green: decay rate (dark) and frequency (light) of the observed oscillations. Dashed line: analytical expression for the oscillation frequency [cf. Eq.~\eqref{eq:freq_modulations}]. (b) Derivative of the eigenvalue associated to the eigenvalue with smallest real part (blue) with respect to the sensing parameter $\Omega_\mathrm{d}$. Derivative of the eigenvalue associated to the oscillations in the WTD (green) with respect to the sensing parameter $\Omega_\mathrm{d}$. Parameters: $\Omega_\mathrm{e}=0.2\gamma_\mathrm{g}$, $\Omega_\mathrm{d}/\Omega_\mathrm{e}=0.01$, $\Omega_{\mathrm{u}} = \Omega_{\mathrm{e}}$, $\Delta_\mathrm{u}/\gamma=0$ (Setup 2) and $\eta_\mathrm{u}=0$. }
 \label{fig:4LS_spectral_analysis}
\end{figure}
In the following section, we will firstly discuss differences between $W'(\tau)$ ($\eta_\mathrm{g}=\eta_\mathrm{u}=1$) and the waiting time distribution $W(\tau)$ ($\eta_\mathrm{g}=1,\eta_\mathrm{u}=0$) [cf. Fig.~\ref{fig:4LS_analytical}(b)]. In a second step we study how $W(\tau)$ changes with inefficient detection $\eta_\mathrm{g}\ll1$, see Fig.~\ref{fig:WTD_osc}(b).

Assuming $\eta_{\mathrm{u}} = 0$ allows us to interpret the result of the evolution between two jumps with $\mathcal{L}_0$ as the waiting time distribution, since every time a photon was detected the system is found in state $\ket{g}$. However, the non-hermitian perturbation ansatz is due to the inefficient detection in the second channel not valid anymore and one has to evolve the system according to a master equation [cf. Eq.~\eqref{eq:no_jump_evolution}]. Nevertheless, as the numerical results illustrated in Fig.~\ref{fig:4LS_analytical}(b) show, there is no drastic change in $W'(\tau)$ for $\eta_{\mathrm{u}}\to 0$. In fact, the oscillations, i.e. the feature we are interested in, do not change much. While the frequency of the oscillations is sustained for $\eta_{\mathrm{u}} = 0$, the minima in $W'(\tau)$ [cf. inset of Fig.~\ref{fig:4LS_analytical}(b)] are not going to zero anymore. Physically, the latter means that, without detecting photons from the decay to level $\ket{u}$, times between jumps with previously vanishing probability occur now with a finite probability.

In order to analyze the change in $W'(\tau)$ and $W(\tau)$ more thoroughly we perform a numerical analysis of the eigenvalues associated to the no-jump generator $\mathcal{L}_0$ [cf. Eq.~\eqref{eq:prob_traj}]. To this end, one vectorizes the no-jump evolution given in Eq.~\eqref{eq:no_jump_evolution}. For $\eta_\mathrm{g}=1,\eta_\mathrm{u}=0$ the imaginary part of the eigenvalue associated to the oscillations, i.e. the frequency, corresponds, as illustrated in Fig.~\ref{fig:4LS_spectral_analysis}(a), to the analytical eigenvalue for the case $\eta_\mathrm{g}=\eta_\mathrm{u}=1$ [cf. Eq.~\eqref{eq:freq_modulations}].

If we now consider the case of $\eta_\mathrm{g}\leq1$, we firstly observe that for unit detection efficiency, i.e. $\eta_{\mathrm{g}} = 1$, the smallest decay rate $\mathrm{Re}(\lambda_0)$ is of the same order as the one of the oscillations [cf. Fig.~\ref{fig:4LS_spectral_analysis}(a)]. While for $\eta_{\mathrm{g}} \to 0$ the smallest decay rate decreases, the decay rate associated to the oscillations increases. This separation results in an increased tail in the waiting time distribution that shows no modulation, as illustrated in Fig.~\ref{fig:WTD_osc}(b). Further, the susceptibility of the eigenvalues associated to $\mathcal{L}_0$ with respect to changes in the parameter $\Omega_\mathrm{d}$ can be used to understand the behavior of the FI for decreasing detection efficiency, see Fig.~\ref{fig:sensing_osc}. The oscillatory modes are not only more responsive to variations in the parameter of interest $\Omega_\mathrm{d}$, as indicated by the derivative of their eigenvalues with respect to this parameter, but they also preserve this sensitivity for $\eta_{\mathrm{g}} \ll 1$ in contrast to $\lambda_0$ [cf. Fig.~\ref{fig:4LS_spectral_analysis}(b)]. Thus, the optimal observation time $t'$ given by the maximum of the FI rate for inefficient detection [cf. Fig.~\ref{fig:sensing_osc}] emerges due to the separation of the two timescales in combination with the eigenvalue capturing the decay for long times, $\lambda_0$, being less sensitive to changes in $\Omega_\mathrm{d}$.

\subsection{Dephasing}\label{Sec:osc_deph_app}

\begin{figure}[t!]
 \centering
 \includegraphics[width=\columnwidth]{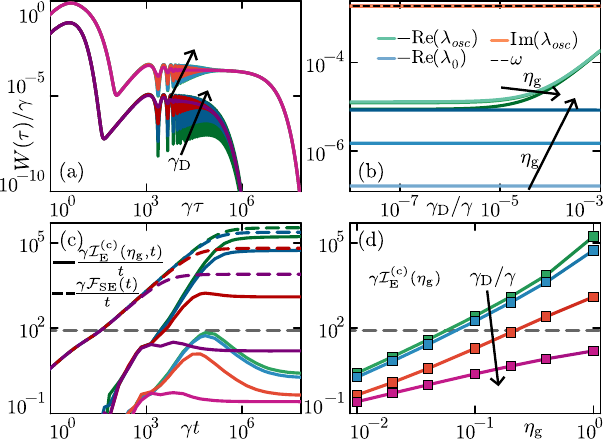}
 \caption{{\bf Impact of dephasing on the sensitivity bounds for Setup 2.} (a) WTD for $\eta_{\mathrm{g}}=1$ (dark) and $\eta_{\mathrm{g}}=0.01$ (light; shifted up by $10^3$ for better representation) and dephasing rates $\gamma_\mathrm{D}/\gamma=0,10^{-5},10^{-4},10^{-3}$ (green, blue, red, purple). (b) Eigenvalues of the no-jump generator $\mathcal{L}_0$ for varying dephasing rate $\gamma_\mathrm{D}$. Overall decay rate (blue), decay rate (green) and frequency (red) of the observed oscillations with $\gamma_{\mathrm{g}}=1,0.1,0.01$ (dark to light). Dashed line: analytical expression for the oscillation frequency [cf. Eq.~\eqref{eq:freq_modulations}]. (c) QFI rate (dashed) and FI rate (solid) for $\eta_\mathrm{g}=0.01, 1$ (light, dark) and dephasing rates $\gamma_\mathrm{D}/\gamma=0,10^{-5},10^{-4},10^{-3}$ (green, blue, red, purple). (d) Asymptotic FI rate for varying detection efficiency $\eta_\mathrm{g}$ and dephasing rates $\gamma_\mathrm{D}/\gamma=0,10^{-5},10^{-4},10^{-3}$ (green, blue, red, purple). The asymptotic QFI rate $\gamma \mathcal{F}_{\mathrm{SE}}/t$ for the corresponding blinking system $\Delta_{\mathrm{u}}/\gamma=0.3$ [cf. Fig.~\ref{fig:FI_blinking}(a)] is for comparison given by the grey dashed line. The FI rate was calculated from a sample of $10^6$ trajectories. Parameters: $\Omega_\mathrm{e}=0.2\gamma_\mathrm{g}$, $\Omega_\mathrm{d}/\Omega_\mathrm{e}=0.01$, $\Omega_{\mathrm{u}} = \Omega_{\mathrm{e}}$ and $\Delta_\mathrm{u}/\gamma=0$ (Setup 2).}
 \label{fig:4LS_dephasing}
\end{figure}

In this section, we  present and discuss the impact of dephasing to the enhanced sensitivity observed in the case of Setup 2.

The results derived in the context of the perturbation ansatz show that the oscillations emerging on an intermediate timescale rely on coherences between the dark state $\ket{D}$ and the $\ket{d}$-level. Thus, the latter and consequently also the concomitant enhanced sensitivity are  susceptible to processes harmful to these coherences. For dephasing of the $\ket{d}$-level at a rate $\gamma_\mathrm{D}$ we indeed find, as illustrated in Fig.~\ref{fig:4LS_dephasing}(a), strong damping of the modulations in the WTD. This behavior can again be understood by an analysis of the eigenvalues associated to the no-jump generator $\mathcal{L}_0$, which shows that the frequency of the oscillations and the overall decay rate are robust to dephasing, see Fig.~\ref{fig:4LS_dephasing}(b). In contrast, the decay rate associated to the oscillations quickly increases for $\gamma_\mathrm{D}/\gamma>\num{1e-5}$ [cf. Fig.~\ref{fig:4LS_dephasing}(b)]. Since the previously observed enhancement in the sensitivity relies on these oscillations, the QFI rate as well as the FI rate decrease for increasing dephasing rates. Nevertheless, for intermediate dephasing rates up to $\gamma_\mathrm{D}/\gamma = 10^{-5}$ we can, as illustrated in Fig.~\ref{fig:4LS_dephasing}(c), recover the original FI rate to a large extent. The asymptotic FI rate depicted in Fig.~\ref{fig:4LS_dephasing}(d) further shows the overall dependence of the sensitivity on the dephasing rate, independent of the detection efficiency.

\section{Optimality of photon counting}\label{app:optimality_photon_counting}

\begin{figure}[t!]
 \centering
 \includegraphics[width=\columnwidth]{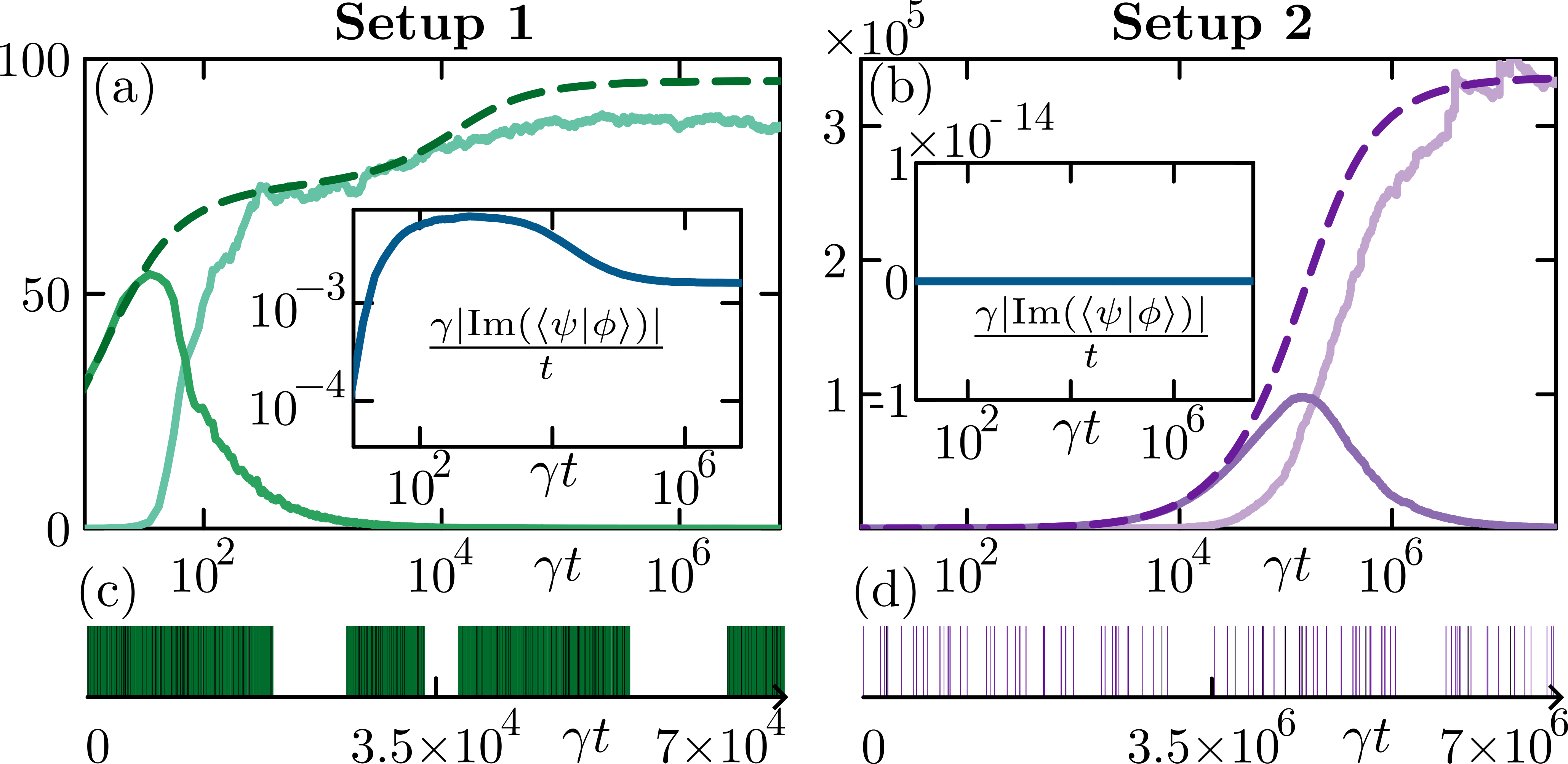}
 \caption{{\bf Optimality of photon counting.} System-environment QFI $\gamma\mathcal{F}_{\mathrm{SE}}/t$ (dashed), conditional QFI of the system $\gamma \mathcal{F}_{\mathrm{S}}^{(c)}/t$ (solid-dark) and classical FI $\gamma \mathcal{I}_\mathrm{E}^{\mathrm{(c)}}/t$ (solid-bright) averaged over $10^4$ trajectories for (a) the blinking system (Setup 1, $\Delta_\mathrm{u}/\gamma=0.3$) and (b) the metastable dark state system (Setup 2, $\Delta_\mathrm{u}/\gamma=0$). In the inset we show $\gamma |\mathrm{Im}(\langle \psi|\phi\rangle)|/t$ , which is a quantity entering in the saturation conditions given in Eq.~(8) of Ref.~\cite{mattes2025designing}. Typical emission record for photon counting with detection efficiency $\eta_{\mathrm{g}}=\eta_{\mathrm{u}}=1$ for the blinking (c) and metastable dark state system (d). In addition to emissions from the $\ket{e}-\ket{g}$ transition, indicated by the green and purple vertical lines for Setups 1 and 2, respectively, we also observe emissions from the $\ket{e}-\ket{u}$ transition (black). Parameters: $\Omega_\mathrm{e}=\Omega_\mathrm{u}=0.2\gamma_\mathrm{g}$, $\Omega_\mathrm{d}/\Omega_\mathrm{e}=0.01$ and $\eta_{\mathrm{g}}=\eta_{\mathrm{u}}=1$.}
 \label{fig:optimality_condition}
\end{figure} 

In this section, we discuss, within the framework developed in Ref.~\cite{mattes2025designing}, if photon counting as a classical measurement is optimal. 

\begin{figure}[t!]
 \centering
 \includegraphics[width=\columnwidth]{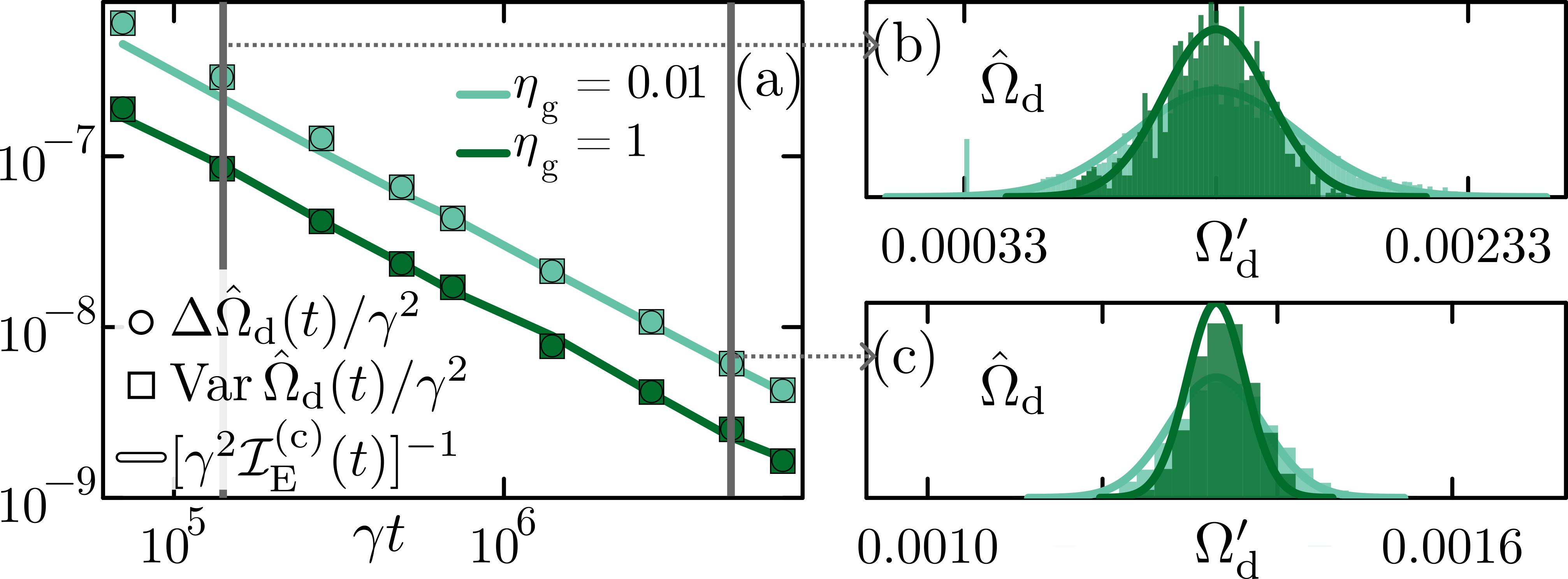}
 \caption{{\bf Maximum likelihood estimator for Setup 1.} (a) Evolution of the variance of the MLE $\mathrm{Var}\,\hat{\Omega}_\mathrm{d}(t)/\gamma^2$  (square), MSE $\Delta\hat{\Omega}_\mathrm{d}(t)/\gamma^2$ (circle) and the inverse of the FI $[\gamma^2 \mathcal{I}_\mathrm{E}^{\mathrm{(c)}}(\Omega_\mathrm{d},t)]^{-1}$ (solid line) for one trajectory of length $\gamma t$ with detection efficiency $\eta_{\mathrm{g}}=1$ (dark) and $\eta_{\mathrm{g}}=0.01$ (light). The statistical analysis used $\num{1e3}$ and $\num{1e4}$ trajectories, respectively. Distribution of the estimator $\hat{\Omega}_\mathrm{d}$ for $\eta_{\mathrm{g}}=0.01$ (light) and $\eta_{\mathrm{g}}=1$ (dark) and trajectories of length  (b) $\gamma t=\num{1.4e5}$  and (c) $\gamma t=\num{4.9e6}$. The lines correspond to normal distributions with mean equal to the true value and variance given by the inverse of the FI. Parameters: $\Omega_\mathrm{e}=0.2\gamma_\mathrm{g}$, $\Omega_\mathrm{d}/\Omega_\mathrm{e}=0.01$ and $\Delta_\mathrm{u}/\gamma=0.3$ (Setup 1).}
 \label{fig:MLE_blinking_app}
\end{figure}

Optimality is in this context identified by the saturation of the system-environment QFI $\mathcal{F}_{\mathrm{SE}}$ [cf. Eq.~\eqref{eq:Fisher_deformed}] with the photon counting QFI $\mathcal{F}_{\mathrm{SE}}^{(c)}=\mathcal{F}_{\mathrm{S}}^{(c)}+\mathcal{I}_\mathrm{E}^{\mathrm{(c)}}$.  The first term of the latter expression captures the information about the parameter stored in the conditional state of the system, given by the conditional QFI of the system $\mathcal{F}_{\mathrm{S}}^{(c)}$~\cite{albarelli2018restoring,mattes2025designing}. The second term corresponds to the classical FI associated to the emission record $\bold{i'_t}=\{t_1^{L_1}, \dots, t_N^{L_N}\}$ involving now also emissions from the transition $\ket{e}\leftrightarrow\ket{u}$, i.e. $\eta_\mathrm{g}=\eta_\mathrm{u}=1$, where $L_j\in\{L_\mathrm{g}, L_\mathrm{u}\}$ with $j=1,\dots,N$. Notice that for long measurement times, the contribution of the FI associated to the emission record dominates over the conditional QFI, as the former increases linearly with time~\cite{albarelli2018restoring,mattes2025designing} [cf. Fig.~\ref{fig:optimality_condition}]. Thus, one has to consider the jump operators $L_\mathrm{g}=\sqrt{\gamma_\mathrm{g}} \ket{g}\!\bra{e}$ and $L_\mathrm{u}=\sqrt{\gamma_\mathrm{u}} \ket{u}\!\bra{e}$ in the photon counting unravelling~\cite{Wiseman2009}
\begin{equation}\label{eq:unrav}
    \begin{split}
        &K_0 = \mathbf{1}- iH\delta t - \frac{\gamma}{2} \ket{e}\!\bra{e}\delta t\, ,\\\
        &K_1^{(\mathrm{g})} = \sqrt{\gamma_\mathrm{g} \delta t} \ket{g}\!\bra{e}\, ,\quad K_1^{(\mathrm{u})} = \sqrt{\gamma_\mathrm{u} \delta t} \ket{u}\!\bra{e}\, ,
    \end{split}
\end{equation}
where we used that $\gamma = \gamma_\mathrm{g}  + \gamma_\mathrm{u}$.

Applying the saturation conditions given in Eq.~(8) of Ref.~\cite{mattes2025designing} to the model given in Eq.~\eqref{eq:Hamiltonian} one finds that the class of saturating states is spanned by states of the form
\begin{equation}
    \ket{\psi_{\mathbf{i_t'}}}=A\ket{g}+B\ket{e}+C\ket{d}+D\ket{u}\, ,
\end{equation}
with $A,D\in\mathbf{I}$ and $B,C\in\mathbf{R}$ (or $A,D\in\mathbf{R}$ and $B,C\in\mathbf{I}$). Here, $\ket{\psi_{\mathbf{i'_t}}}$ denotes the state conditioned on a specific emission record $\bold{i'_t}$. For the free evolution between two jumps defined in Eq.~\eqref{eq:unrav}, we find with the model [cf. Eq.~\eqref{eq:Hamiltonian}]
\begin{equation}
K_0=\left(\begin{array}{cccc}
     1&-i\Omega_\mathrm{e}\delta t&-i\Omega_\mathrm{d}\delta t& 0 \\
     -i\Omega_\mathrm{e}\delta t&1-\gamma \delta t/2&0&-i\Omega_\mathrm{u}\delta t\\
     -i\Omega_\mathrm{d}\delta t&0&1&0\\
     0&-i\Omega_\mathrm{u}\delta t&0&1-i\Delta_\mathrm{u}\delta t
\end{array}\right)\, .
\end{equation}
Therefore, one remains within the class of saturating states for $\Delta_\mathrm{u}=0$, which corresponds to the system featuring metastable dark states (Setup 2).

As illustrated in Fig.~\ref{fig:optimality_condition}, while the saturation conditions [cf. Eq.~(8) of Ref.~\cite{mattes2025designing}] are for $\Delta_\mathrm{u}=0$ fulfilled and the FI rate approaches the fundamental limit given by the QFI rate, for $\Delta_\mathrm{u}\neq0$ the saturation conditions are not fulfilled and the FI rate does not saturate the QFI rate. Nevertheless, we note that for the blinking system (Setup 1) and unit efficiency in both channels $\eta_{\mathrm{g}}=\eta_{\mathrm{e}}=1$ we asymptotically find $\mathcal{I}_\mathrm{E}^{\mathrm{(c)}}/\mathcal{F}_{\mathrm{SE}}=0.9$. Furthermore, by comparing the emission pattern of Setup 1 given in Fig.~\ref{fig:optimality_condition}(c) with the one of Setup 2 in Fig.~\ref{fig:optimality_condition}(d) one immediately sees the influence of the dark state leading in the latter case to a reduced emission rate and an emission pattern displaying no intermittency.

Although for the four-level scheme with detuning, i.e. featuring blinking, photon counting is strictly not optimal, we recall that its simpler three-level version studied in appendix~\ref{app:3LS} also satisfies the optimality conditions of Ref. \cite{mattes2025designing}. Thus, it constitutes an example of optimality with classical measurements robust to inefficiency.

\section{Maximum likelihood estimator}\label{sec:MLE_app}

In this section, we provide additional results for the maximum likelihood estimator. In Fig.~\ref{fig:MLE_blinking_app} we show additional results on the MLE and the FI in the context of asymptotic optimality for detection efficiencies $\eta_{\mathrm{g}}=1, 0.01$ and the blinking system (Setup 1).

\bibliography{refs.bib}
\end{document}